\documentclass[prd, aps, superscriptaddress, preprintnumbers, floatfix, nofootinbib]{revtex4}

\pdfoutput=1

\usepackage{enumitem}
\usepackage{braket}
\usepackage{hyperref}
\usepackage{tablefootnote}
\usepackage{amsmath,amssymb,amsfonts,mathrsfs}
\usepackage{booktabs}
\usepackage{mathtools}
\usepackage{bm}
\usepackage{dcolumn}
\usepackage{graphicx}   
\usepackage[latin1]{inputenc}
\usepackage{latexsym}
\usepackage{color}

\newcommand\be{\begin{equation}}
\newcommand\ba{\begin{eqnarray}}
\newcommand\ee{\end{equation}}
\newcommand\ea{\end{eqnarray}}

\begin{document}

\title {Extracting the Signal of Cosmic String Wakes from 21-cm Observations}

\author{David Maibach}
\email{maibach@student.ethz.ch}
\affiliation{Physics Department, ETH, CH-8093 Z\H{u}rich, Switzerland}

\author{Robert Brandenberger}
\email{rhb@physics.mcgill.ca}
\affiliation{Department of Physics, McGill University, Montr\'{e}al, QC, H3A 2T8, Canada} 

\author{Devin Crichton}
\email{dcrichton@ethz.ch}
\affiliation{Institute for Particle Physics and Astrophysics, ETH, Wolfgang-Pauli-Strasse 27, CH-8093 Z\H{u}rich, Switzerland} 

\author{Alexandre Refregier}
\email{alexandre.refregier@phys.ethz.ch}
\affiliation{Institute for Particle Physics and Astrophysics, ETH, Wolfgang-Pauli-Strasse 27, CH-8093 Z\H{u}rich, Switzerland} 

\date{\today}


\begin{abstract}

A cosmic string wake produces a distinct non-Gaussian signal in 21-cm intensity maps at redshifts above that of reionization. While the string signal is (locally) larger in amplitude than the signal of the Gaussian fluctuations of the $\Lambda$CDM model, they are overwhelmed (even locally in position space) by astrophysical and instrumental foregrounds. Here, we study to what extent the signal can be extracted from noisy interferometric data. The narrowness of the string-induced feature in redshift direction allows for a subtraction of astrophysical and instrumental foregrounds. Based on the specific geometry of the string signal we identify a particular three-point statistic which is promising in order to extract the signal, and we find that, having in mind a telescope of specifications similar to that of the MWA instrument, the string signal can be successfully extracted for a value of the string tension of $G\mu = 3 \times 10^{-7}$. Prospects for further improvements of the analysis are discussed.

\end{abstract}



\maketitle


\section{Introduction} 
\label{sec:intro}

A large class of particle physics models beyond the Standard Model admit topologically stable string configurations (see \cite{CSrevs} for review articles on cosmic strings). If Nature is described by such a theory, then a network of strings will form during a symmetry breaking phase transition in the early universe and persist, by causality \footnote{By causality, the direction in field space in which the field condensate falls cannot be correlated on length scales larger than the horizon, thus leaving behind a network of imperfections in the condensate, the cosmic strings.}, until the present time \cite{Kibble}. These strings are relativistic, i.e. characterized by a tension whose magnitude is equal to the energy per unit length. Any curvature of such a string will hence induce relativistic motion.The network of strings takes on a {\it scaling solution} in which the statistical properties of the string distribution are independent of time if all lengths are scaled to the Hubble radius. Strings carry trapped energy, and their gravitational effects lead to specific non-Gaussian signatures in a wide range of cosmological observables (see e.g. \cite{RHBrev1}). The search for these signatures provides an interesting interplay between particle physics and cosmology \cite{RHBrev2}. Note that since the energy per unit length $\mu$ of a cosmic string is of the order $\eta^2$, where $\eta$ is the scale of the particle physics symmetry breaking, searching for signals of cosmic strings can constrain particle physics from the high energy end downwards, in contrast to accelerator experiments which constrain new physics from bottom up. Finding a cosmic string signal would be a spectacular new discovery and would provide new directions for particle physics model building. The absence of string signals would yield new upper bounds on the energy scale $\eta$ of symmetry breaking.

The network of cosmic strings consists of a random-walk-like distribution of ``infinite'' (or ``long'') strings, and a distribution of string loops with radii $R$ smaller than the Hubble radius. The mean curvature radius of the infinite string network is of the order of the Hubble radius. This is maintained in time by the long strings intersecting and chopping off string loops \cite{CSrevs}. A long string will typically undergo one intercommutation event per Hubble time. Hence, we can model \cite{onescale} the distribution of long strings at time $t$ as a set of string segments of length $c_1 t$ which live for a Hubble time. There are $N$ string segments per Hubble volume. Here, $c_1$ and $N$ are constants of the order one which must be determined from numerical string evolution simulations \cite{CSsimuls}. The centers and orientations of the string segments can be taken to be random, and uncorrelated in different Hubble time steps.

Space perpendicular to a cosmic string is conical with deficit angle $8 \pi G \mu$, where $G$ is Newton's gravitational constant \cite{conical}. The extent of the deficit angle extends to one Hubble distance away from the string \cite{Joao}. This feature leads to the two key effects of an ``infinite'' string, the first being the lensing of photons passing on different sides of a moving string. This induces a line discontinuity in the temperature of cosmic microwave background (CMB) photons \cite{KS} of magnitude $8 \pi G \mu v_s \gamma_s$, where $v_s$ is the velocity of the string and $\gamma_s$ is the corresponding relativistic $\gamma$ factor. The signature in CMB temperature maps is a rectangle in a CMB map of comoving size $c_1 t_i \times v_s \gamma_s t_i$, where $t_i$ is the time when the string segment in present. The Canny algorithm \cite{Canny}, wavelet and shapelet statistics \cite{Hergt} and machine learning techniques \cite{Razvan} have been proposed in order to search for this signal.

Because of the conical deficit angle, a long string segment moving through the primordial gas will lead to an overdensity in its wake.  The comoving planar dimensions of this overdensity (called a ``wake'' \cite{wake}) are set by the length and distance travelled by the segment, and the thickness is set by the deficit angle. The key point is that wakes are nonlinear overdense regions which exist at arbitrarily early times after the phase transition. In particular, at any time after $t_{rec}$ they lead to regions of enhanced baryon number and  thus of enhanced free electron density. The extra Thomson scattering in wakes thus leads to a characteristic CMB polarization signal \cite{Holder1}, a rectangle in the sky with extra polarization (including a B-mode component). The overdensity of baryons inside of a string wake also leads to the signal which we will focus on in this paper, namely a wedge of extra 21-cm absorption or emission in 21-cm redshift maps \cite{Holder2}.

Since string wakes are disrupted by the Gaussian fluctuations from the $\Lambda$CDM model which are the dominant source of structure formation \footnote{For string tensions consistent with the current upper bound \cite{CMBbound} $G \mu < 1.5 \times 10^{-7}$ coming from measurements of the angular power spectrum of the CMB temperature maps, cosmic strings cannot be the dominant source of structure formation.} at redshifts smaller than that of reionization (which we denote by $z_{EoR}$) \cite{Disrael}, the 21-cm signal of strings wakes is most clearly visible for $z > z_{EoR}$. In fact, at these redshifts the wakes are the dominant source of nonlinear fluctuations in the Standard Model. Hence, in this paper we study the potential of planned interferometric 21-cm redshift surveys to discover the signal of a string wake given the multitude of astrophysical and instrumental foregrounds \footnote{Note that cosmic strings also lead to a global 21-cm signal (see \cite{Oscar} for a study of the global 21cm signal of string wakes, and \cite{us1, us2} for a study of the corresponding signal for superconducting cosmic string loops).}.

In the following section, we review the 21-cm signal of a cosmic string wake. In Section 3 we study the string wake signal in Fourier space in which interferometers will acquire the data. Section 4 describes our numerical simulations. The 21-cm maps which we produce include the signal of a string wake, the signals of the primordial Gaussian fluctuations plus astrophysical foregrounds and instrumental effects. We describe the foregrounds which we include and explain how we model instrumental effects. We focus on two statistics which we use to differentiate between maps with and without strings: a $\chi^2$ statistic and a three point function with a shape designed to pick out the string signal. We also describe signal processing techniques which we use to suppress the backgrounds relative to the string signal. Our results are presented in Section 5. We find that an experiment like MWA has the angular resolution and sensitivity to clearly identify the cosmic string signal using our three point statistic provided we apply our signal processing techniques to suppress the foregrounds. This is true even if a small patch of the sky of size \(5^\circ \times 5^\circ\)  is analyzed. The benchmark value of the string tension which we use is $G\mu = 3 \times 10^{-7}$. While this value is a factor of 2 larger than the current upper bound, we show that strings with tensions comparable and slightly lower than this bound can also be identified, provided that a larger patch of the sky is analyzed.

A few words concerning units and notation. We use natural units in which the speed of light $c$, Boltzmann's constant and the Planck constant are set to one. In these units $G \mu$ is dimensionless. We will be working in the context of a spatially flat Friedmann-Lemaitre-Robertson-Walker metric with physical time $t$, comoving spatial coordinates ${\bf{x}}$ and a scale factor $a(t)$ normalized to be one at the present time $t_0$. The Hubble expansion rate is given by $H(t) \equiv {\dot{a}}/a$, and its inverse is the Hubble radius. Relevant times for our study are $t_{rec}$, the time of recombination, and $t_{eq}$, the time of equal matter and radiation. As usual, the cosmological redshift at time $t$ is denoted by $z(t)$.

\section{Cosmic String Wakes} \label{wakes}

\subsection{Setup}

In this section we study in more detail the distinct effects that influence the signal of a string wake in a 21-cm map. The goal is to acquire intuition for the appearance of the string wake in interferometer data in order to implement realistic simulations and apply appropriate analysis techniques. We focus on an understanding of the size, the shape and the brightness temperature  of the string wake and transfer the results to Fourier space. As interferometer surveys sample measurement data in  the form of frequency modes, the latter step is essential if we want to compare with real survey data.

We will first focus on the calculation of the brightness temperature intensity of the string wake signal in position space. There are two major processes that affect the temperature of the atoms inside the string wake and consequently its brightness temperature, shock heating and diffusion.   We then transform our signal to Fourier space and stress critical implications for the detectability. We also address the extent of the string signal in redshift space and how it influences the Fourier space data for a fixed redshift hypersurface.

As already mentioned in the Introduction (\ref{sec:intro}), a single cosmic string segment moving through the primordial gas generates a region of twice the background density of matter in its wake.  The comoving planar dimensions of this wake \cite{wake}) are set by the length and distance travelled by the segment. The width of a wake increases linearly from zero (at the front of the wake, the position of the string at the end time) to the value $8 \pi v_s \gamma_s G \mu t_i$ at the back of the wake. The initial overdensity creates a gravitational force on matter above and below the wake, and this causes the comoving width of the wake \footnote{Defined as the distance from the central plane of the wake of the matter shell which is decoupling from the Hubble flow and starting to collapse onto the wake.} to grow in time as $(z(t) + 1)^{-1}$ (see e.g. \cite{Zeld}).  Thus, the comoving dimensions of a wake created by a string segment laid down at time $t_i$ and viewed at time $t$ are
\be
c_i t_1 \, \times \, v_s \gamma_s t_i \, 
\times \, 4 \pi G \mu v_s \gamma_s t_i \frac{z(t_i) + 1}{z(t) + 1} \, .
\ee 
The geometry of a wake in three dimensional physical space is sketched in Fig. 1. The horizontal axis corresponds to the plane perpendicular to our line of sight, the vertical axis is the direction in the line of sight. The string segment was moving from the right to the left. The induced wake is thickest at the original position of the string, and minimal at the final position.  
\begin{figure}[!tb]
\centering
\includegraphics[scale=1.3]{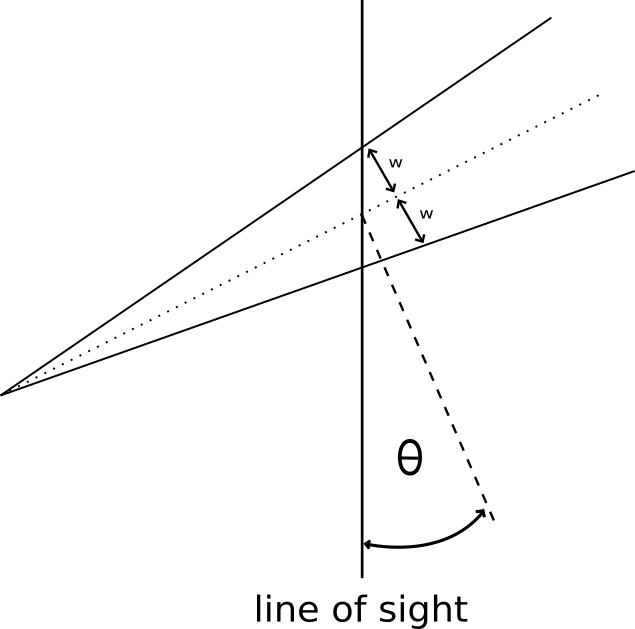}
\caption{Sketch of a string wake in three-dimensional real space (dimensions not scaled) being tilted relative to the line of sight. The width of the wake, as it was defined above is marked relative to the center of the wake, here displayed as a dotted line. The dashed line represents a vector normal to the wake plane and defines the angle \(\theta\).}
\label{fig:10}
\end{figure}

We are interested in the density distribution of baryons inside the string wake. However, since for times $t < t_{rec}$ baryons are still coupled to radiation we first focus on the clustering of cold dark matter due to cosmic strings (studied e.g. in \cite{coldDM}). This is a reasonable approach as for $t > t_{rec}$ baryons will rapidly fall into the potentials created by cold dark matter clustering. Since cosmic strings produced at earlier times are more numerous, we focus on wakes generated at redshifts of around $z_{eq}$. Note that before $t_{eq}$ there is no gravitational clustering of cold dark matter and consequently no wake growth.

The overdensity of cold dark matter will attract baryons and, in particular, hydrogen atoms which remain neutral after the background temperature drops below the ionisation threshold until the onset of reionization, when ionizing sources form in the universe. During this period, overdense regions of hydrogen can emit or absorb radiation of wavelength \(\lambda=21\) cm, corresponding to the hyperfine transition of electron in the ground state of the hydrogen atom (HI). As a result, the string wake in this period will be visible either via absorbing \(21\) cm background radiation or emitting extra intensity at this frequency. 

First, we test the type of signature we expect the string to produce in the dark ages.
The variable that determines weather a cosmic string absorbs or emits extra \(21\) cm radiation is the temperature of the atoms inside the string wake. Therefore, we calculate the kinetic temperature of the HI atoms inside the wake. We closely follow the corresponding calculation in \cite{Holder2}.

In analogy to hydrodynamics, when the in-falling matter shells that collapse towards the center of the wake collide with other streams of matter, a shock occurs. This appears at approximately half the maximal distance of the matter shells with respect to the center of the wake \cite{Sornborger_1997}. Since the distance of turnaround is half the width of the shell without any perturbations, the shock occurs at approximately one quarter of the unperturbed matter shell width. Equivalent to what was stated before, this implies that the averaged density inside the shocked region is four times the background density. We define the distance of the shocked region above the plane of the initial overdensity as the width of the string wake.

\subsection{Shock-Heated Wakes} \label{sec_theo_cosmicstring}

In order to compute the brightness temperature induced by a string wake, we need to determine the kinetic temperature of HI atoms inside the wake. In a first analysis we neglect the initial thermal velocities of the gas particles falling forwards the center of the wake. The temperature is then determined by the process of shock heating. To find this temperature we calculate the physical velocity of the gas particles at the point where the shock occurs, and then infer the resulting temperature at redshift $z$ via the standard formula \(\frac{3}{2}k_bT=\frac{1}{2}mv^2\). Here, \(m\) is the mass of HI atoms and \(v\) the velocity of the attracted particles. Consequently, it follows that \cite{Holder2}
\begin{align}\label{SW9}
    T_K \simeq [20 \text{K}](G\mu\cdot 10^6)^2 (\gamma_sv_s)^2\frac{z_i+1}{z+1}.
\end{align}
This implies that wakes produced at earlier times exhibit higher temperatures due to shock heating. The overdense region grows as more and more matter has time to collapse onto it, yielding a wedge-like shape as discussed earlier. 

Note that a string wake created at some redshift $z_i$ persists until the present time, growing in comoving thickness as described above. Our past light cone can intersect such a wake at any redshift $z < z_i$, and we will see the signal of the wake from this redshift which we call the {\it redshift of emission}. We need to compare \(T_K\) with \(T_\gamma = 2.725 (1+z)\) K, the CMB temperature. For \(T_K<T_\gamma\) the wake displays an absorption signal, otherwise it establishes a region of extra emission. The result will depend on the chosen values for string speed \(v_s\), tension \(\mu\) and redshifts of emission of radiation \(z\) and generation of the string \(z_i\) 

The result for the kinetic temperature of HI atoms inside the wake is based on the assumption that they thermalize solely due to shock heating effects. We discuss corrections to the results which result from including the primordial kinetic energy of the gas in the next subsection. First, we apply the description of \(T_K\) to a cosmological context and calculate how the signal appears to an observer on earth when competing against the background CMB \(21\) cm radiation. 

Determining the brightness temperature of cosmic string wakes, we apply the equation of radiative transfer along the line of sight piercing this overdensity. For \(21\) cm emission, the brightness temperature \(T_b(\nu)\) at an observed frequency \(\nu\) reads 
\begin{align}\label{SW5}
    T_b(\nu) \, = \, T_S(1-e^{-\tau_\nu})+T_\gamma(\nu)e^{-\tau_\nu},
\end{align}
where \(T_S\) is the spin temperature and \(T_\gamma\) the CMB temperature. \(\tau_\nu\) corresponds to the optical depth which is given by the integral of the absorption coefficient along the path of the photons through the wake, \(\tau_\nu=\int \alpha ds\). Note that the first term in Eq. \eqref{SW5} represents spontaneous emission, while the second term is due to absorption and stimulated emission. We do not give a detailed description of all processes involved in radiative transfer, and we refer to \cite{Furlanetto_2006} for more detailed descriptions of 21 cm emission at high redshift. 

We are interested in the comparison of the emission from the HI clouds created by the wakes with the \(21\) cm radiation originating from the CMB. Therefore, we consider
\begin{align}\label{SW6}
    \delta T_b(\nu)\, = \, \frac{T_b(\nu)-T_\gamma(\nu)}{1+z}\approx \frac{T_S-T_\gamma(\nu)}{1+z}\tau_\nu,
\end{align}
where in the first equation we inserted Eq. \eqref{SW5} and expanded the exponential to first order in \(\tau_\nu\), and where \(z\) is the redshift of emission. Note that in this formula we assumed that except for the wake the line of sight crosses no other density perturbations.  It is also worth emphasising, that for redshift \(<200\) the brightness temperature from a unperturbed region in space is negative as the atoms in this region exhibit a lower temperature than the one of CMB photons at that time. The spin temperature \(T_S\) is defined via 
\be
n_1/n_0 \, = \, 3\exp(-T_\star/T_S) \, , 
\ee
where \(n_1\) and \(n_0\) are the number densities of atoms in the two hyperfine energy states and \(T_\star = E_{10}/k_B=0.068\) K is the temperature corresponding to the energy difference between these two states. Taking into account collisions of HI atoms as well as scattering of UV photons, the spin temperature is given by
\begin{align}
    1-\frac{T_\gamma}{T_S}= \frac{x_c}{1+x_c+x_\alpha}\big(1-\frac{T_\gamma}{T_K}\big)+\frac{x_\alpha}{1+x_c+x_\alpha}\big(1-\frac{T_\gamma}{T_C}\big),
\end{align}
which boils down to
\begin{align}\label{SW8}
    T_S \, = \, \frac{1+x_c+x_\alpha}{1+(x_c+x_\alpha)\frac{T_\gamma}{T_K}}T_\gamma,
\end{align}
when reionization is not yet significant so that \(T_C=T_K\) \cite{Furlanetto_2006}. Here, \(T_C\) is the color temperature and the coupling coefficients for collision \(x_c\) and for UV scattering \(x_\alpha\) are given by the scattering rates of hydrogen atoms with electrons and UV photons respectively. In this work we will neglect \(x_\alpha\) \footnote{We comment on implications of a non-vanishing \(x_\alpha\) after Eq. \eqref{SW11}.} and consider only \(x_c\).

The optical depth (see e.g.  \cite{Furlanetto_2006}) of a hydrogen cloud is given by
\begin{align}
    \tau_\nu \, = \, \frac{3c^2A_{10}}{4\nu^2}\frac{\hbar\nu_{10}}{k_bT_S}\frac{N_{HI}}{4}\phi(\nu),
\end{align}
with the column density \(N_{HI}\) of HI and the line profile \(\phi(\nu)\) which is normalized such that \(\int \phi(\nu) d\nu=1\). \(N_{HI}\) is given by the number density of hydrogen atoms inside the wake \(n_{HI}^{wake}\) integrated over the length the light ray traverses the cloud. This length depends on the width of the wake and its alignment relative to the line of sight. Let us define the angle \(\theta\) as the angle enclosed by the line of sight and a vector normal to the plane marking the center of the wake, see Figure \ref{fig:10}. Then,
\be
N_{HI} \, = \, \frac{2n_{HI}^{wake} w}{\cos \theta} \, ,
\ee
where the factor \(2\) arises since the distance \(w\) measures the width with respect to the center of the wake. 
\begin{figure}[!tb]
\centering
\includegraphics[scale=1.3]{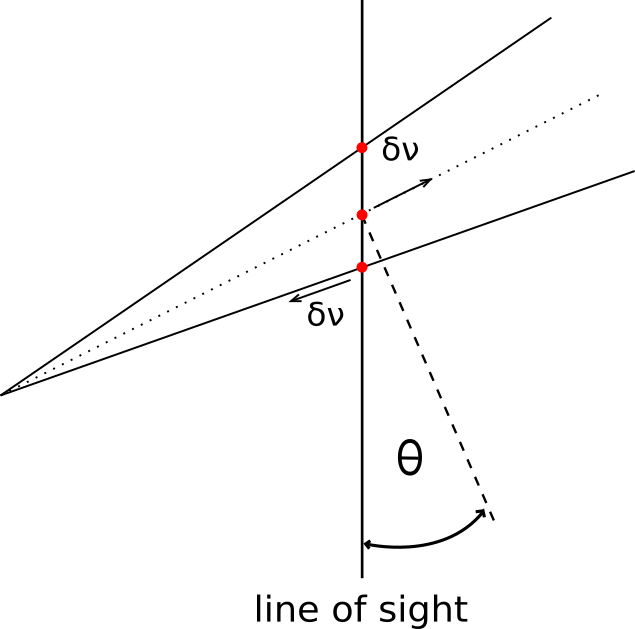}
\caption{Sketch of photons emitted from different points inside the wake (red dots) in three-dimensional real space. The three points along the line of sight mark points of emission of \(21\) cm radiation from the bottom, the center and the top of the wake. Due to Hubble expansion of the wake in planar dimensions, the lower and upper points experience a Doppler shift in frequency of the \(21\) cm photons, \(\delta \nu\), relative to the point at the center of wake plane. Note that the sketch is not to scale. The angle \(\theta\) is equivalent to the definition in Figure \ref{fig:10}.}
\label{fig:11}
\end{figure}

The line profile \(\phi(\nu)\) describes  the broadening of the emission line and is generally influenced by bulk motion, thermal motion and pressure effects. The thermal motion in our case is negligible as are pressure effects since we make the standard astrophysical assumption of small pressure. The critical factor for the line profile in our consideration is the bulk motion. The wake is broadened in the planar directions due to cosmic expansion. For a non-trivial angle \(\theta\) this leads to an effective difference for the frequencies of \(21\) cm photons reaching the observer from the top and the bottom of the wake \cite{Holder2} (for an illustration see Figure \ref{fig:11}). If we use the photons emitted at the center of the wake as a reference, then the photons at the highest point and at the lowest point of the wake experience a relative Doppler shift due to the expansion of
\begin{align}\label{SW17}
    \frac{\delta \nu}{\nu} \, = \, 2 \sin(\theta)\tan(\theta)\frac{H(z)w(z)}{c},    
\end{align}
where \(H\) is the Hubble expansion rate and \(w\) the width of the wake, both evaluated at the redshift of emission.\footnote{The frequency shift \(\delta \nu\) is the essential factor determining the extent of the wake in redshift space (see Section (\ref{sec_filters_simulations}).} Inserting both quantities and with \(c=1\) we find
\begin{align}
    \frac{\delta \nu}{\nu} \, = \, \frac{48\pi}{15}G\mu\gamma_s v_s\sin(\theta)\tan(\theta) (z_i+1)^{1/2}(z+1)^{-1/2} \, .
\end{align}
The angle \(\theta\) is displayed in Figures \ref{fig:10} and \ref{fig:11}. With an appropriate normalization, we find
\begin{align}\label{SW7}
    \phi(\nu) \, = \, \frac{1}{\delta \nu},
\end{align}
for \(\nu\in[\nu_{10}-\frac{\delta \nu}{2},\nu_{10}+\frac{\delta \nu}{2}]\), otherwise \(\phi(\nu)=0\). Now that we determined every parameter in Eq. \eqref{SW6} for the string wake, we insert \(T_S\) in Eq. \eqref{SW8} and the optical depth with Eq. \eqref{SW7} as a line profile and find 
 \begin{align}
     \delta T_b(\nu) \, = \, 2\frac{x_c}{1+x_c}\bigg(1-\frac{T_\gamma}{T_K}\bigg)\frac{3c^3A_{10}\hbar}{16\nu_{10}^2k_BH_0\Omega_m^{1/2}}n^{bg}_{HI}(t_0)\frac{n_{HI}^{wake}(t_0)}{n^{bg}_{HI}(t_0)}(2\sin^2(\theta))^{-1}(1+z)^{1/2} \, ,\notag
\end{align}
where \(\Omega_m\) is the fraction of the energy density in matter. As we stated before, the ratio of the HI number density inside the wake and background number density is \(n_{HI}^{wake}(t_0)/n^{bg}_{HI}(t_0)=4\) for a shock heated wake. The Hubble constant was rescaled to its current value using \(H(z)=H_0\Omega_m^{1/2}(1+z)^{3/2}\). The width of the wake has explicitly canceled out but implicitly, it still influences the wake temperature \(T_K\) and the signal's appearance in redshift space. Finally, with \(A_{10}=2.80\cdot 10^{-15}\text{s}^{-1}\), \(T_\star=0.068\) K, \(H_0=73 \text{km}\text{ s}^{-1}\text{Mpc}^{-1}\), \(\nu_{10}=1420\) MHz, \(\Omega_b=0.042\), \(\Omega_m=0.26\) we obtain
\begin{align}\label{SW11}
    \delta T_b(\nu) \, = \, [0.07\text{ K}]\frac{x_c}{1+x_c}\bigg(1-\frac{T_\gamma}{T_K}\bigg)(1+z)^{1/2}(2\sin^2(\theta))^{-1}.
\end{align}
As mentioned earlier, including UV scattering corresponds to the substitution \(x_c\rightarrow x_c+x_\alpha\) in the above equation. Since the factor \(x/(1+x)\) is monotonically increasing and \(x_\alpha>0\), this yields a larger relative brightness temperature \(\delta T_b\) compared to the case of \(x_\alpha=0\). Neglecting UV scattering in this calculation hence corresponds to a conservative estimate of \(\delta T_b\).

Note that the collision coefficient \(x_c\) is governed by the de-excitation cross section \(\kappa_{10}^{HH}\) and reads \cite{Zygelman_2005}
\begin{align}
    x_c \, = \, \frac{n\kappa_{10}^{HH}T_\star}{A_{10}T_\gamma}.
\end{align}
For the exact values of \(\kappa_{10}^{HH}\) we use Table 2 in \cite{Zygelman_2005}.  A linear interpolation between the points in that table is too imprecise. Therefore, we numerically fit a smooth function of \(T_K\) to the provided data points. As a result, \(\kappa_{10}^{HH}\) can be described by
\begin{align}\label{joder}
    \kappa^{HH}_{10} \, = \, 10^{-10}\bigg[27.6 - \frac{27.59}{\big[1 + \big(\frac{T_K}{24.03}\big)^{2.28}\big]^{0.014}}\bigg].
\end{align}
The improvement in accuracy through the application of the fit function becomes evident in section \ref{subsec_optdect_simulataions}. Note that the factor \((2\sin^2(\theta))^{-1}\) in Eq. \eqref{SW11} does not lead to a physical divergence. Any actual measurement of the brightness temperature is carried out with finite frequency resolution. Integrating over the frequency interval, the factor of \((\sin \theta)^{-1}\) cancels out.

We add a few comments on the assumptions we made during the calculation. First note that for certain values of string wake characteristics, i.e. combinations of alignment (expressed through \(\theta\)), speed \(v_s\) and tension \(\mu\) the signal switches form absorption to emission. By solving \(T_K=T_\gamma\) with respect to the string tension, one finds the value
\begin{align}
    (G\mu\times 10^6)^2 \, \simeq \, 0.1 (\gamma_s v_s)^{-2}\frac{(1+z)^2}{1+z_i},
\end{align}
for the critical tension at the point at which the signal changes form absorption to emission \cite{Holder2}. 

Second, the approximation of neglecting the initial thermal energy of the gas particles breaks down when the temperature inside the wake is smaller than the temperature of the background gas when it is adiabatically compressed due to the overdensity of the wake. Consequently, \(T_K\) as it was given in Eq. \eqref{SW9} does no longer hold in this case. The cosmic gas temperature as a function of redshift is given by 
\begin{align}
    T_g \, = \, 0.02 \text{ K} (1+z)^2 ,
\end{align}
when Compton heating through the CMB is neglected \cite{Seager}. For adiabatic compression of a mono-atomic gas to \(n_{compressed}/n_{uncompressed}=4\) the temperature gains a factor of \(4 ^{2/3}\approx 2.5\). Hence, the breakdown condition reads \(T_K=2.5T_g\). Above \(2.5T_g\) the temperature of the HI atoms inside the wake is well approximated by \(T_K\) from Eq. \eqref{SW9}. Below this threshold, the initial gas temperature effects dominate. Shock heating then becomes subdominant. In the following subsection, we will discuss the implications for the brightness temperature of wakes with subdominant shock heating.

\subsection{Diffuse Wakes}

In the previous analysis, we neglected the intrinsic temperature of the gas at \(z_i\) when the wake is laid down. We mentioned that gravitational accretion onto the cosmic string wake increases the background gas temperature by a factor of roughly \(2.5\), boosting the temperature due to thermal motion up to \(2.5T_g\)  where \(T_g\) is the background gas temperature. As stated before, one can ignore the incoherent thermal motion of the accreted gas for \(T_K>2.5T_g\), where \(T_K\) is the temperature of atoms thermalized via shock heating. However, when the incoherent velocities due to thermal motion dominate over the coherent velocity of the gas due to gravitational attraction towards the center of the wake, i.e. \(T_K<2.5T_g\), no shock heating occurs. In this case, the overdense region of gas induced by the wake is larger than it is in the case of subdominant thermal motion. The resulting cosmic string wake is called "diffuse wake" \cite{Diffusion}. In this section, we calculate the broadening and the brightness temperature of a diffuse wake following \cite{Diffusion}.

The difference between shock heated and diffuse wakes lies in the density and the spatial extent of the wake. While both types accrete the same amount of mass in linear perturbation theory, the diffuse wake is larger and hence less dense. Let us first consider the thickness of diffuse wakes. For equipartition of energy between potential and thermal energy we find for \(T_K\ll T_g\)
\begin{align}\label{SW10}
    m_{HI}\delta\Phi =\frac{3}{2}T_g,
\end{align}
where \(m_{HI}\) is the mass of a HI atom and \(\delta \Phi\) the gravitational potential of the overdensity, \(\delta \Phi = 2\pi G\sigma |h|\). Here,  \(\sigma\) is the surface density and \(h\) is the height. Inserting the induced gravitational potential in Eq. \eqref{SW10} we obtain a linear scaling of \(h\) with temperature. Translating \(h\) to the width \(w\) of the wake, we find that the width increases by a factor 
\begin{align}
    w(z)|_{T_K<T_g} \, = \, w(z)|_{T_g=0}\cdot \frac{T_g}{T_K},
\end{align}
for \(T_K<T_g\). In the previous section, we saw that the width influences the extent of the wake in redshift space. Consequently, diffuse wakes exhibit a larger extent of the signal in redshift direction. On the other hand, due to the reduced baryon density inside the wake the brightness temperature decreases relative to the shock heated wake. The overdensity of a diffuse wake is thus given by 
\begin{align}
    \Delta \rho (z) \, = \, \frac{\sigma(z)}{w(z)}=\rho_0\bigg(\frac{T_K}{T_g}\bigg),
\end{align}
and hence
\begin{align}
    \frac{\rho}{\rho_0} \, = \, \bigg(1+ \frac{T_K}{T_g}\bigg).
\end{align}
Since \(T_K\) decreases as we move to higher redshift but \(T_g\) increases quadratically, the overdensity will be significantly smaller at earlier times yielding less \(21\) cm absorption or emission.

Using this result, we can now determine a formula for the string brightness temperature of a diffuse wake. At first sight, the expressions for the wake brightness temperature for a shock heated and diffuse wake are very similar. For a shock heated wake we have Eq. \eqref{SW11},
\begin{align}\label{SW12}
    \delta T_b(z) \, = \, [17 \text{ mK}]\frac{x_c}{1+x_c}\bigg(1-\frac{T_\gamma}{T_K}\bigg)\frac{n_{HI}^{wake}}{n_{HI}^{bg}}\frac{(1+z)^{\frac{1}{2}}}{2\sin^2 \theta},
\end{align}
where \(z\) corresponds to the redshift of \(21\) cm emission or absorption and the other entities are defined as in Eq. \eqref{SW11}. The latter equation holds for \(T_K>2.5T_g\). For convenience, we extend its validity up until \(T_K>3T_g\). Then, for \(T_K<3T_g\), the density of HI atoms changes according to the calculation above, i.e. we replace \(n_{HI}^{wake}/n_{HI}^{bg}\) with \((1+T_K/T_g)\). In addition, if \(T_g\) exceeds \(T_K=3T_g\) the temperature inside the string wake is governed by the background gas temperature including the adiabatic compression factor, yielding a brightness temperature of
\begin{align}\label{SW13}
    \delta T_b(z) \, = \, [17 \text{ mK}]\frac{x_c}{1+x_c}\bigg(1-\frac{T_\gamma}{3T_g}\bigg)\bigg(1+ \frac{T_K}{T_g}\bigg)\frac{(1+z)^{\frac{1}{2}}}{2\sin^2 \theta}.
\end{align}
The relative brightness temperature of a string wake follows the latter equation for \(T_K<3T_g\).

Let us summarize the findings of this section: The cosmic string in relativistic motion produces a wedge like overdensity in three-dimensional real space. The brightness temperature of this overdensity when compared to the CMB \(21\) cm radiation is generally described by 
\begin{align}\label{SW14}
    \delta T_b(z) \, = \, [17 \text{ mK}]\frac{x_c}{1+x_c}\bigg(1-\frac{T_\gamma}{T_{K/g}}\bigg)\frac{n_{HI}^{wake}}{n_{HI}^{bg}}\frac{(1+z)^{\frac{1}{2}}}{2\sin^2 \theta}.
\end{align}
The governing temperature that determines the amplitude and the sign of the latter, i.e. if the signal is visible in emission or absorption, \(T_{K/g}\), is given by 
\begin{align}\label{SW15}
    T_{K/g} \, = \, \begin{cases} 
      T_K & T_K>3T_g \\
      3T_g & T_K\leq3T_g \\
   \end{cases},
\end{align}
where in the first case shock heating dominates the thermalization of the HI atoms inside the wake created by the string while in the second case the incoherent intrinsic gas velocity due to thermal motion determines the wake temperature. A similar piece-wise definition holds for the density fraction in Eq. \eqref{SW14}:
\begin{align}\label{SW16}
     \frac{n_{HI}^{wake}}{n_{HI}^{bg}} \, = \, \begin{cases} 
      4 & T_K>3T_g \\
      \bigg(1+ \frac{T_K}{T_g}\bigg) & T_K\leq3T_g \\
   \end{cases}.
\end{align}
In addition, diffusion effects influence the width of the wake as 
\begin{align}\label{SW17}
    w(z) \, = \, \begin{cases} 
      w(z) & T_K\geq T_g \\
      w(z)\cdot \frac{T_g}{T_K} & T_K<T_g \\
   \end{cases}.
\end{align}
Equations \eqref{SW14} - \eqref{SW17} provide us with a complete description of the wake's brightness temperature, density and width for relevant domains of redshift. These formulas will be used in the simulations which will be described later.  There, we will also discuss the transition point between shock-heated and diffuse wakes in more detail, as it plays an important role in the investigation of the string detectability.

\section{Cosmic String Signal in Interferometric Data} \label{signal}

 In this work, we investigate the signature of a cosmic string wake in interferometer data. Since interferometers probe the three-dimensional matter distribution via two angular and one frequency dimension, characteristics of the wake's appearance in all these dimensions need to be analysed. The goal of this section is to explorate the string wake's unique features that distinguish it from Gaussian noise as well as from other non-Gaussian signals in interferometer data. This knowledge can then be applied later to filter out the comparably faint string wake signal from noise contaminated interferometer data.

We briefly mentioned earlier that the string wake can take multiple alignments with respect to the angular plane. As we will see in this section, these alignments affect the spatial extent in real and Fourier space of the signature we expect to observe as well as its length in redshift direction. 

\subsection{Wake alignment in redshift directions}\label{subsec_fourier_cosmicstring}

In general, an interferometer probes a patch of the sky with a given frequency resolution, i.e. it detects the intensity of the radiation of this patch within a band \(\Delta \nu\). For modern day interferometers, this band varies between \(\mathcal{O}(10^1)\) and \(\mathcal{O}(10^2)\) kHz. If the sky patch is small enough, we can apply the flat sky approximation and parametrize the patch via two Euclidean coordinates. Measuring multiple frequency bins and combining them yields a three-dimensional measurement of this patch where the extent in the third dimension, the frequency direction, is given by the collection of the frequency bins considered. Assuming a cosmic string wake is located within this probed region, its signature might extend over multiple frequency/redshift bins.\footnote{Note that we use the terms frequency and redshift space interchangeably during this section as they are equivalent to each other up to some factor of proportionality.} In redshift space, the string wake has a distinctive shape. Due to the line broadening illustrated in section \ref{sec_theo_cosmicstring} the signature will appear as a wedge-like region of \(21\) cm absorption or emission. It is wide in angular directions and narrow in redshift direction. For most alignments relative to the angular plane, the wake signal is wide only in one angular direction if we consider a fixed redshift. As the interferometer cannot distinguish between frequencies inside one bin, in the measurement data of a single frequency bin the segment of the wake lying inside this bin appears to be projected onto the angular plane. Ergo, it is the Fourier signal of the projected shape of \(21\) cm emission or absorption that the interferometer picks up and that we aim to filter out of the data.

\begin{figure}[!tbp]
  \centering
  \begin{minipage}[b]{0.45\textwidth}
    \includegraphics[width=\textwidth]{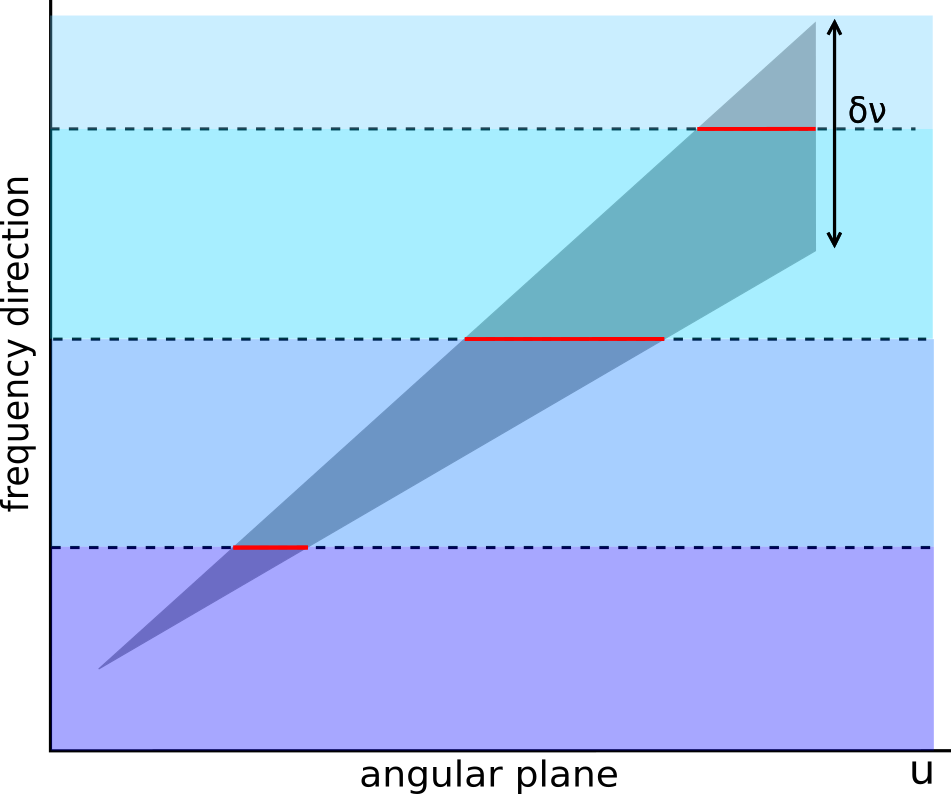}
    \caption{Sketch of the geometry of the wake in redshift space. The horizontal axis represents one coordinate of the angular plane (the x-y plane), the vertical axis the frequency direction. The overdense wake is darkly shadowed and the frequency direction is separated into differently coloured redshift bins. The angular intervals where the wake crosses the boundary between two redshift bins are marked in red in order to facilitate comparison with Fig. \ref{fig:15}. The geometry of the wake is not to scale.}
    \label{fig:14}
  \end{minipage}
  \hfill
  \begin{minipage}[b]{0.45\textwidth}
    \includegraphics[width=\textwidth]{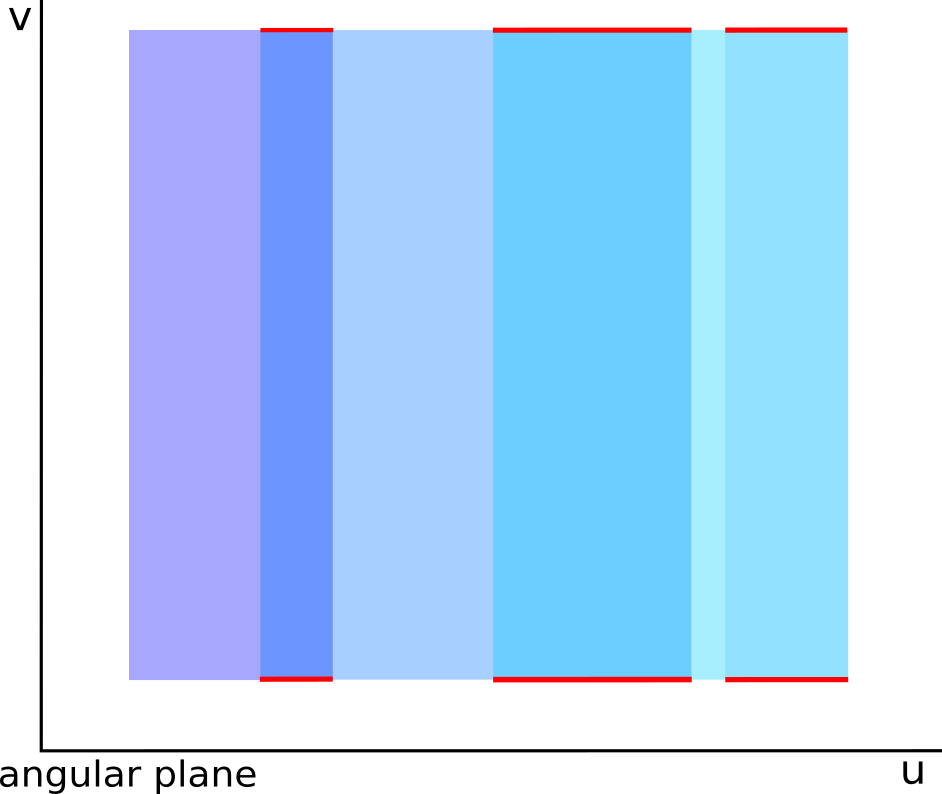}
    \caption{Sketch of the projection of the wake with orientation given in \ref{fig:14} onto the angular plane. The string signals in the redshift bins of \ref{fig:14} are indicated with the corresponding colours. Some parts of the wake yield a signal in neighbouring redshift bins, and these regions are marked in red. 
    \\\\\\ }
    \label{fig:15}
  \end{minipage}
\end{figure}

The alignment of the wake is determined by the plane spanned by the tangent vector to the (straight) string segment and its direction of motion.  In redshift space, the direction of the wake is given by its orientation relative to the light cone of the observer. For a wake moving away form the observer on earth, its tip will intersect the past light cone of the observer earlier than its back, yielding a larger redshift compared to the back of the wake. Moreover, the width of the wake in redshift direction vanishes at its tip and is maximal at its back. This results from its wedge like shape in real space and the consequent line broadening \(\delta \nu\) (see Figures \ref{fig:11} and \ref{fig:14} and Eq. \eqref{SW17}) due to Hubble expansion in its angular dimensions. All in all, the geometry of the wake in frequency direction is captured by Figure \ref{fig:14}. For a more detailed derivation of this geometry we refer to \cite{Holder2}. 

With the given shape in redshift direction, the binning results in splitting of the signal into separate parts. Combining the redshift bins in the sense of projecting them onto the \(u,v\)-plane we find the shape resembling the complete projection of the wake onto this plane. The process is illustrated in Figures \ref{fig:14} and \ref{fig:15}. In Figure \ref{fig:14} we chose a random alignment for the string wake to describe the binning process.  We colored each redshift bin and marked the regions in the angular plane where the projections will overlap. For each bin, we project the wake segment onto the plane. In Figure \ref{fig:15}, we patched all bins together and thus, in some sense, projected the whole wake onto the plane. Note that each segment is colored corresponding to the redshift bin it lies in.  In the simplest case of string alignment, the string wake lies entirely in one redshift bin, yielding the largest extent of the wake in the x-y plane. Since \(\delta \nu\) (see Eq. \eqref{SW17} and Figure \ref{fig:14}) is in general smaller than the instrumental redshift resolution mentioned above, this special alignment is a valid simplification.  However, even in the case of a general alignment the wake does not extend over many redshift bins compared to other sources of \(21\) cm radiation. Note that this in an important feature of the string wake signal that allows us to separate it from other Gaussian and non-Gaussian signals which the interferometer picks up. 

Another point worth emphasizing is that as a consequence of the wedge-like shape of the string wake in redshift space, its projection onto the angular plane leads to a gradient in the intensity. The tip of the wake exhibits the smallest intensity, the back of the wake the maximal value as we integrated the signal over its extent in redshift space \(\delta z\) when projecting. In this work we will not make use of this feature.

So far we have not discussed one important aspect of the wakes alignment, namely that its projected shape varies with the alignment. The wake in general has a rectangular shape in its wide dimensions. Whether we see a nice rectangular shape in the observer's x-y plane depends on the orientation of the wake relative to our line of sight. For simplicity, we will consider alignments in which the area of the projection of the total wake is \(\geq1/4\) of the wakes actual area. 

In conclusion, the alignment of the wake influences its projected shape onto the angular plane and, therefore, its appearance in measured data. In this analysis, we restrict ourselves to a selected range of possible alignments favorable for a detection of a string wake. The fact that the string wake signal is narrow in redshift direction will allow us to extract the signal from maps including noise, since most types of noise contaminants that are relevant for the \(21\) cm frequency domain are correlated over many redshift bins, while the string signal covers comparatively few. This distinction allows for the application of noise removal techniques which increase the strings detectability. Note that in general, the frequency resolution of the interferometer is a crucial criterion to select suitable interferometer survey for potential string detection. If redshift bins are too large, the noise will dominate within the bin and consequently the detectability declines.
 
\subsection{Wakes in Fourier space}\label{ch_cosmicstring}

In the previous section, we analyzed the distinctive shape of the string signal in position-redshift space and outlined unique features of the wake that can be used advantageously for a potential detection.  As an interferometer receives the signal data in the form of frequency modes, we now focus on distinct features of the string signal in Fourier space. Our goal is the identification of unique features that can be used to construct wake-specific filter methods and statistics. 

In the following we assume that the wake is located inside a single redshift bin and therefore its projected shape onto the angular plane is rectangular with edge lengths which we here denote by \(X\) and \(Y\). The Fourier transform of any function \(f(\mathbf{x})\) over a two-dimensional space \(\mathbf{A}\) with \(\mathbf{x}=(x, y)\) is  
\begin{align}
    F_f(\mathbf{k}) \, = \, \int_\mathbf{A}f(\mathbf{x})e^{-2\pi i \mathbf{k}\cdot \mathbf{x}}\text{d}\mathbf{x}
\end{align}
where \(\mathbf{k}\) is the spatial frequency vector \(\mathbf{k}=(k_x,k_y)\) - chosen so that the linear scale corresponding to \(k\) is \(2\pi/k\). The function \(f\) in our case describes the rectangular string signal inside the coordinate plane. We start with the simplest possible alignment, namely a rectangle whose edges are parallel the the coordinate axis. The center of the rectangle resides at point \((a,b)\). Then, 
\begin{align}\label{FS1}
    f(x,y) \, = \, \begin{cases} 
                                  1 & \text{if $a\leq x \leq X+a \And b\leq y \leq Y+b$} \\
                                  0 & \text{else} \\
  \end{cases},
\end{align}
and the Fourier transform is given by
\ba
   F(k_x,k_y) \, &=& \, \int \int f(x,y)e^{-i2\pi(k_xx+k_yy)}dxdy \nonumber \\
   &=& \, \int_{a}^{X+a}e^{-i2\pi k_xx}dx\int_{b}^{Y+b}e^{-i2\pi k_yy}dy \nonumber \\
    &=& \, \frac{1}{\pi k_x}\sin{(\pi k_xX)}e^{-i\pi k_x(2a+X)}\frac{1}{\pi k_y}\sin{(\pi k_y Y)}e^{-i\pi k_y(2b+Y)}.\label{FS2}
\ea

\begin{figure}[!tbp]
  \centering
  \begin{minipage}[b]{0.45\textwidth}
    \includegraphics[width=\textwidth]{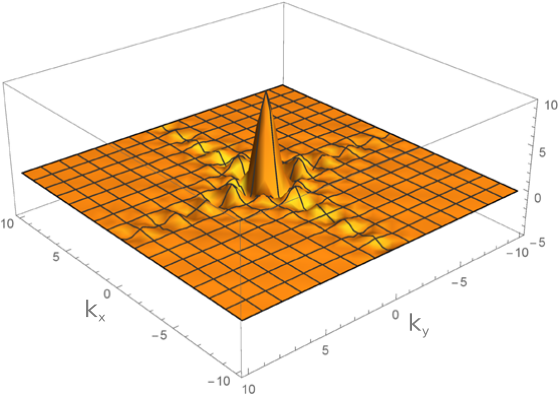}
    \caption{Fourier transform \(F(k_x,k_y)\) for \(X=Y=1.5\) in the case of \(a=-\frac{X}{2}\) and \(b=-\frac{Y}{2}\).}
    \label{fig:16}
  \end{minipage}
  \hfill
  \begin{minipage}[b]{0.45\textwidth}
    \includegraphics[width=\textwidth]{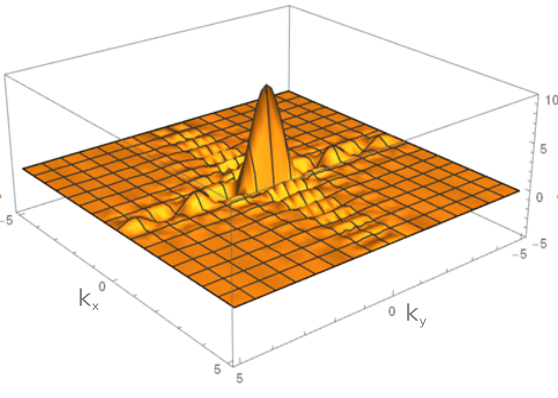}
    \caption{Fourier transform \(F(k_x,k_y)\) for \(X=1.5\), \(Y=1\) in the case of \(a=-\frac{X}{2}\) and \(b=-\frac{Y}{2}\).}
    \label{fig:17}
  \end{minipage}
\end{figure}

We immediately see that any displacement of the center of the wake away from the coordinate origin of our \(x,y\)-plane yields an imaginary part so that the Fourier transform becomes complex. For \(a = -X/2\) and \(b=-Y/2\), i.e. the square center equals coordinate origin, the Fourier transform is purely real. Note also that the signal is damped in each direction as \(\sim 1/k_{x/y}\). In case of \(F\in \mathbb{R}\), the length of the rectangular projection of the string wake enters solely in the \(\sin\)-function of Eq. \eqref{FS2}. It follows that for smaller side lengths the sine function has a longer periodicity in the corresponding \(k\)-coordinate, and we expect a stretching in this direction. At the origin, the signal converges as \(\sin(\alpha k)/k \rightarrow \alpha\) for \(k\rightarrow 0\). We illustrate the features of \(F(k_x,k_y)\) in Figures \ref{fig:16} and \ref{fig:17}. In Figure \ref{fig:16} the side lengths \(X\), \(Y\) are equal, in Figure \ref{fig:17} \(Y\) is smaller than \(X\) resulting in a stretching in \(k_y\)-direction. 

To obtain the amplitude of the wake signal in Fourier space, we insert as our function $f$ the relative brightness temperature \(\delta T(x, y)\) at the redshifted \(21\) cm line. This way, the Fourier transform effectively yields \(\delta T(k_x,k_y)\) which gives the intensity in \(k\)-space. As an interferometer measures the intensity of a certain \(k\)-mode domain for a pixel in the sky given by the angular resolution, \(\delta T(k_x,k_y)\) yields an accurate portrait of the signal in interferometer data.

\begin{figure}[!tbp]
  \centering
  \begin{minipage}[b]{0.45\textwidth}
    \includegraphics[width=\textwidth]{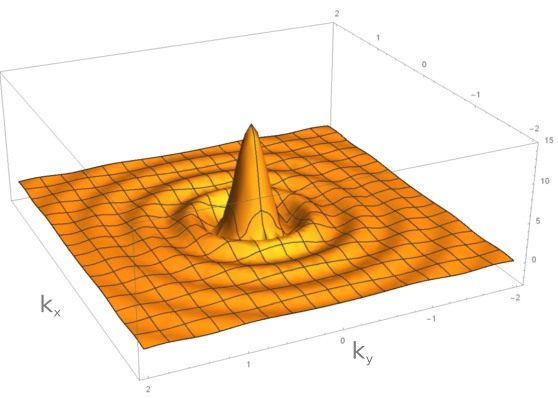}
    \caption{Fourier transform of the projection of a spherical clump overdensity onto the angular plane. We transform according to Eq. \eqref{FS1}.\\}
    \label{fig:18}
  \end{minipage}
  \hfill
  \begin{minipage}[b]{0.45\textwidth}
    \includegraphics[width=\textwidth]{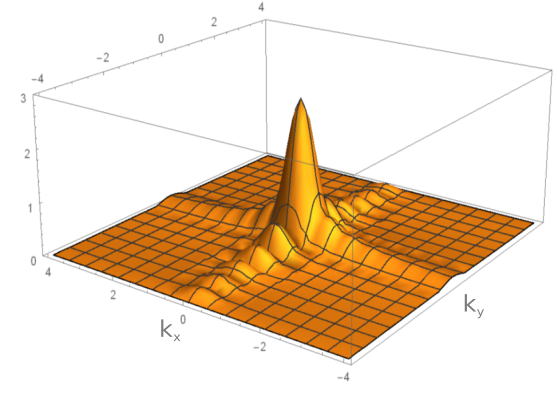}
    \caption{Absolute value of the Fourier transform \(F(k_x,k_y)\) for \(X=1.5\), \(Y=1.5\) and \(a=-\frac{X}{2}\), \(b=-\frac{Y}{2}\) where we introduced an intensity gradient in \(x\)-direction of the real space projected wake area.}
    \label{fig:19}
  \end{minipage}
\end{figure}

The first features that can be unambiguously connected to the string wake signal are the four perpendicular ridges in Figures \ref{fig:16} and \ref{fig:17}. These ridges result from the transformation of the sharp edges of the projected wake signal into Fourier space. Comparing the wake signal with other common structures for overdensities, such as spherical clumps (Figure \ref{fig:18}), the ridges emerge as a wake-specific feature.  

As mentioned in the section before, for a wake residing in a single redshift bin, its projection onto the angular plane gains an intensity gradient due to the varying thickness in redshift direction \(\delta z\). Assuming the wake is wedge shaped and oriented in the \(x\)-direction, its projection yields the highest intensity for the \(x\)-coordinate at the back of the wake. In direction tangent to the string segment, i.e. for \(x=\text{const.}\), the intensity is constant. The \(x\)-dependence of the wake is thus described by 
\begin{align}
\delta z \cdot \delta T_b \cdot \bigg|\frac{x-x_{tip}}{x_{back}-x_{tip}}\bigg|
\end{align}
for \(x\in [x_{back},x_{tip}]\) and else \(=0\), where \(\delta z\) is the wake thickness in redshift direction, \(x_{back}\) is the \(x\)-coordinate of its back and \(x_{tip}\) of the wakes tip. Plugging this result into the Fourier transformation, the ridge in \(x\)-direction becomes smooth. The signal also becomes complex so that we have to take its absolute for appropriate visualization. Note that taking the absolute gives an intuition for the power spectrum $P$ of the string wake as \(P(k_x,k_y)\sim |F(k_x,k_y)|^2\)  
In Figure \ref{fig:19} we display the absolute value of the Fourier transformed wake signal including an intensity gradient. Instead of the ridges with oscillations in amplitude in \(k_x\) direction for \(k_y\approx 0\) we find an almost smooth damping \(\sim 1/k_x\). Except for this smoothing the wake signal does not change significantly and the geometries are the same. Thus, in the following we will ignore the intensity gradient of the wake and focus instead on the alignment of the ridges.

\begin{figure}[!tbp]
  \centering
  \begin{minipage}[b]{0.45\textwidth}
    \includegraphics[width=\textwidth]{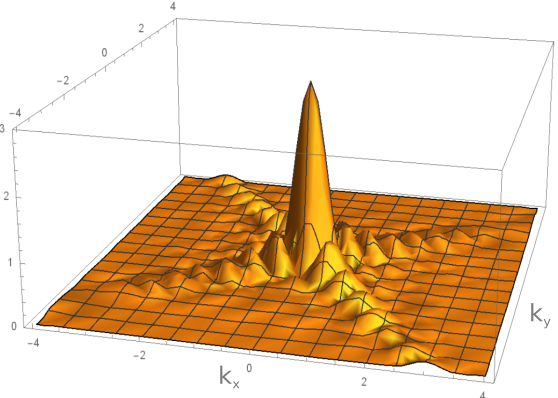}
    \caption{Absolute value of the Fourier transform \(F(k_x,k_y)\) for \(X=1.5\), \(Y=2\). The rectangular wake projection is rotated by \(\alpha \approx 37^\circ\) relative to the coordinate cross in real space. The intensity over the signal area is homogeneous.}
    \label{fig:20}
  \end{minipage}
  \hfill
  \begin{minipage}[b]{0.45\textwidth}
    \includegraphics[width=\textwidth]{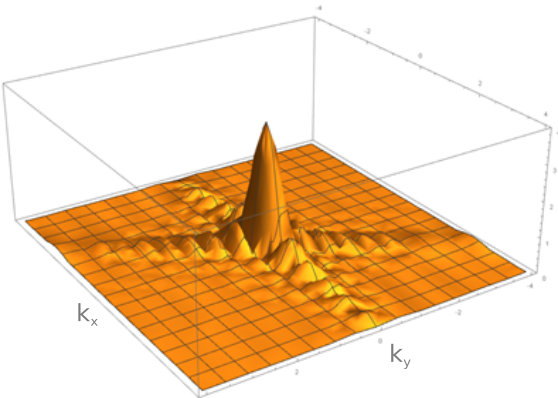}
    \caption{Absolute value of the Fourier transform of a parallelogram with one pair of edges being parallel to the \(x\)-axis, the other two edges enclose an angle of \(\alpha \approx 70^\circ\) with the \(x\)-axis. \\}
    \label{fig:21}
  \end{minipage}
\end{figure}

Let us now consider rotations in the angular plane of a wake centered at the coordinate origin, i.e. \(a = -X/2\), \(b=-Y/2\) in \ref{FS1}. As in the case of a shifted wake, rotations in real space add phases to the Fourier transform.  In Figure \ref{fig:20} we plot the absolute value of the Fourier transform and find exactly the same shape as for the absolute value of the non-rotated wake signal, except for a rotation by an angle corresponding to the angle applied in the two-dimensional angular plane.  
The phase information of the Fourier transform of the rotated wake is highly non-trivial., but we do not make use of this in the following. 

For completeness, let us assume the wake does not reside exactly in one redshift bin but crosses it. Then, the shape of the resulting signal inside this bin projected onto the angular plane differs slightly from what we discussed so far. Note however, that in this case there are still straight edges resulting from the sharply defined projection area of the wake signal intensity against the dark background. The alignment of these edges may change depending on how the wake crosses the redshift bin of interest. A simple example is shown in Figure \ref{fig:21}. Here, we assumed that the projected area of the string inside the redshift bin has the shape of a parallelogram with two edges parallel to the \(x\)-axis and the other two being slightly tilted with respect to the \(y\)-axis. Fourier transforming this configuration results in one pair of ridges being tilted as well in Fourier space. The angles enclosed by two neighboring edges of the parallelogram correspond to the angles enclosed by the ridges. Note also that for a single edge of a rectangle being tilted, Figure \ref{fig:21} would correspondingly show only one ridge rotated by the same angle. Again, this illustrates how the ridges result from the Fourier transform of the straight edges enclosing the signal of intensity \(\delta T_b\) compared to the background with \(\delta T_b =0\). 

In general, an arbitrary alignment of the wake and its extent over multiple redshift bins produces a variety of polygons as projected signal shapes in the \(x,y\)-plane. The important features that distinguish these structures from other overdensities are the ridge-like arms that spread out from the peak at the origin of the coordinate system in Fourier space. The number of ridges, the angles between them and their amplitude (smoothly or periodically damped) depend on the specific alignment of the wake in redshift direction. Note that the alignment of the ridges is in direct correspondence to the alignment of the edges of the projected two-dimensional real space signal. 

In this work, we only consider string wakes residing in a single redshift bin. This simplifies the Fourier signal significantly since the projection onto the angular plane in this case is almost perfectly rectangular. Hence, we are dealing with Fourier signals of the shape displayed in Figure \ref{fig:16}. In addition, we neglect the intensity gradient induced by the thickness of the wake in redshift space, \(\delta z\) (in analogy to Eq. \eqref{SW17}). 

As Figure \ref{fig:16} is the targeted structure in Fourier space resulting from a string wake inside a redshift bin, its features are essential to identify it in interferometer data.\footnote{Figure \ref{fig:16} is from now on referred to as the string wake signal in Fourier space.} The ridges are a distinct element of the wake signal in Fourier space. Since spherical clumps (or other contaminants) also produce a distinct peak at the origin of Fourier space (shown in Figure \ref{fig:18}), we must rely on the ridges as detection characteristic. Choosing a suitable statistic to extract this characteristic is thus essential for being able to detect the signal of cosmic string wakes.

In the following section we use the above insights about the Fourier signal of a wake in order to simulate a realistic string wake signal received by a modern day interferometer. In particular, we apply statistics that enhance the signature of the wake (based on \ref{fig:16}).  

\section{Numerical Simulations}

Our goal is to study if cosmic string signatures are extractable from realistic interferometer data. In the previous sections, we discussed how an overdensity induced by a cosmic string wake appears in data sampled by radio interferometry. We calculated the amplitude of the intensity of the string signature through the induced relative brightness temperature and determined the expected shape a string in the frequency domain for a wake that resides in a single redshift bin, the thickness of the bin given by the finite frequency resolution of the interferometer. In this chapter, we will include the effects of contaminants found in a realistic measurement data.  The biggest noise components are different types of galactic as well as extra-galactic foregrounds and instrument specific noise. We also include the effects of the primordial Gaussian fluctuations which are the source of the origin of structure in the universe. Following the analysis of the contaminants, we use our insights on distinct features of noise and signal to find suitable statistics and filters that can be applied to the data set in order to extract the wake signal from the larger foregrounds. We then discuss the implementation of the numerical simulations. At the end of this section (in subsection \ref{sec_implement_simulations}) we comment on the assumptions used in the simulations, the application of special implementation techniques and selected parts of the program. Note that the full numerical work is available at 
\hyperlink{https://github.com/maibachd/simulationthesis}{\text{GitHub}} \cite{Git}. 

When speaking of interferometers, we refer to antenna array interferometers (and not dish interferometers) since this type is used in most experiments designed to study the large scale of the 21-cm sky, one reason being the lower implementation cost, and another the large number of Fourier modes that can be probed. 

\subsection{Foreground contamination}\label{sec_fg_ch_simulations}

We will focus on signals of string wakes from the dark ages. With the onset of reionization at \(z\approx 12\,\,\)\footnote{There are large uncertainties in literature concerning the exact time reionization starts. For the purpose of this investigation, we want to maximize the redshift interval under consideration and hence chose the value of \(z=12\) for the end of the dark ages.} ionizing radiation enters the clouds of neutral hydrogen, leading to the ionization and excitation, and effectively washing out the clean string signature which we have discussed in the previous section. 

Besides the density fluctuations from the dark ages, interferometers pick up radiation from sources formed during later times like stars, Synchrotron radiation and free-free transitions. The intensity of these so called foregrounds is much larger than the signal we are aiming to probe. Galaxies and extra-galactic point sources, in particular, can be extremely bright and dwarf the cosmological HI signal. In the following we consider galactic Synchrotron radiation, extra-galactic point sources, galactic free-free emission and extra-galactic free-free emission.

For the following analysis we assume that there has already been some sort of foreground removal or subtraction technique applied to the measurement data, i.e. we are not considering the raw output of the interferometer but a pre-filtered version. A number of different methods for modelling and subtracting foregrounds have been proposed in \cite{numerics1, numerics2, numerics3, numerics4, numerics5, numerics6, numerics7, numerics8, numerics9}. The residual contamination can be modelled as a sum of Gaussian processes described by angular power spectra \(C_l^i(\nu_1,\nu_2)\)  for our main noise components. We model their combination as
\begin{align}
    C_l(\nu_1,\nu_2)&:=\epsilon_{fg}^2 \sum_i A_i \bigg(\frac{l_{ref}}{l}\bigg)^{\beta_i}\bigg(\frac{\nu_{ref}^2}{\nu_1\nu_2}\bigg)^{\alpha_i}\exp\bigg(\frac{-\log^2(\nu_1/\nu_2)}{2\xi^2}\bigg),\label{porcozio}\\
    &=\epsilon_{fg}^2 \sum_i C_l^i(\nu_1,\nu_2)
\end{align}
where \(l_{ref}\) and \(\nu_{ref}\) are reference values for mode number and frequency, and \(\epsilon_{fg}\) is a foreground removal factor, i.e. for \(\epsilon_{fg}=1\) we consider the full residual foreground contamination. The power spectra are defined via the correlation functions
\begin{align}
    C_l^i(\nu_1,\nu_2)\equiv \braket{a^i_{lm}(\nu_1)a^i_{lm}(\nu_2)},
\end{align}
\cite{numerics3}. 
where the $a_{lm}$ are the coefficients of the expansion of the noise field in spherical harmonics, and the angular brackets denote spatial averaging. Since we consider very thin redshift bins we can take \(\nu_1 \approx\nu_2\). Note that we assume here that the correlations between the contaminants are negligible. 

In the following we will study what fraction of the foregrounds has to be removed before being able to extract the string signatures. We wish to find a statistic for which no such subtraction is required for values of $G\mu$ which are of interest, i.e. for which we can choose $\epsilon_{fg} = 1$. The smaller \(\epsilon_{fg}\) has to be for the string wake signature to become detectable via a given processing technique, the worse the detectability using this technique is. The values for the parameters and the foregrounds that we consider here are given in Table \ref{tab:1} (see \cite{numerics3} for details).
\begin{table}[t!]
    \centering
\begin{tabular}{ccccc} \toprule
   Foreground type & $[\text{mK}^2]$ & $\beta$ & $\alpha$ & $\xi$ \\ \midrule
    Galactic Synchrotron & 1100 & 3.3 & 2.80 & 4.0\\
    Point Sources &57&1.1 &2.07&1.0\\
    Galactic free-free & 0.088 & 3.0 & 2.15 & 35\\
    Extra-galactic free-free & 0.014 & 1.0 & 2.10 & 35\\\bottomrule
\end{tabular}
    \caption{Fiducial foreground \(C_l(\nu_1,\nu_2)\) model parameter extracted from \cite{numerics3} and adapted to the reference values \(l_{ref}=1000\) and \(\nu_{ref}=130 \) MHz.}
    \label{tab:1}
\end{table}

Note that all four foregrounds which we consider produce a larger fluctuation power at each frequency than the cosmic string \(21\) cm signal. It would be impossible to measure \(21\) cm signals if not for the high coherence of the contaminants across frequencies compared to the very short frequency space correlation length of the string signal. We will discuss the coherence lengths of the foregrounds again in section \ref{sec_filters_simulations}. Our choice of residual foregrounds is in accordance with common literature on cosmology with \(21\) cm intensity mapping (see e.g. \cite{Bull_2015}) and provides the basis for modern day techniques of residual foreground removal such as \cite{deep21}. The parameter values for the power spectra in \cite{numerics3} are determined based on real observational measurements. With the power spectra of the foreground components at hand, we can construct their real space image in the two-dimensional angular plane. In this way we can adapt the simulated patch such that it matches real observational data. For the galactic foregrounds, i.e. Galactic Synchrotron and free-free emission we use the modelling techniques of \cite{numerics7} and references therein. We assume these two types of foregrounds to be Gaussian distributed with the above power law in Fourier space. The same is assumed to hold for extra-galactic free-free emission where we base our assumptions on \cite{fg4}. For point sources, the procedure of modeling them in real space is less trivial. A realistic picture of point sources can be obtained by scattering sources of different flux density across the sky via a Poisson distributed random walk. The Poisson distribution of point sources can potentially contaminate higher order correlation functions and is generally more difficult to model as it requires accurate catalog data. Assuming that extremely bright sources have been removed, i.e. we considering point sources up to a certain threshold flux density limit, and that the patch of the sky under observation contains a large number of point sources, a Gaussian distribution is a sufficiently accurate approximation after removing the means. For an accurate modeling of unresolved point sources we refer the reader to \cite{numerics7}. In this work, we approximate this foreground type to be Gaussian and model it in such a way that we match the results of \cite{fg2} and references therein. 

In addition to foregrounds, the measurement of an interferometer is distorted by noise from the atmosphere and other radio background radiation. This type of noise is considered as instrumental noise and will be discussed in subsection \ref{sec_interfer_ch_simulations}. 

\subsection{$\Lambda$CDM Fluctuations}

While the cosmic string signal is dwarfed in amplitude by the foregrounds, it is larger in amplitude than the Gaussian fluctuations from the $\Lambda$CDM model which lead to cosmological structure formation. The induced \(21\) cm brightness temperature fluctuations from the $\Lambda$CDM model are a biased tracer of the matter density field on large scales \cite{large} and described by the power spectrum 
\begin{align}\label{sim2}
    P_{T_b}(k,z) \, = \, \bar T_b(z)^2(b+f\mu^2)^2P_m(k,z),
\end{align}
where \(b\) is the bias factor and \(P_m(k,z)\) the real-space matter power spectrum. Here, the growth rate \(f\) is given by \cite{Linder_2005}
\be
f \, = \, \Omega_M^\gamma(z) 
\, ,
\ee
with \(\gamma\approx0.55\) for a \(\Lambda\)CDM-cosmology. The parameter \(\mu\) is given by \(\mu=k_{\parallel}/k\). The mean brightness temperature is taken to be 
\begin{align}\label{equ_37}
    \bar T_b(z) \, = \, 0.1\bigg(\frac{\Omega_{HI}}{0.33\cdot 10^{-4}}\bigg)\cdot \bigg(\frac{\Omega_M+(1+z)^{-3}\Omega_\Lambda}{0.29}\bigg)^{-1/2}\cdot \bigg(\frac{1+z}{2.5}\bigg)^{1/2}\text{mK},
\end{align}
following \cite{Chang_2008} and with \(\Omega_{HI}b=0.62\cdot 10^{-3}\). In accordance to \cite{Shaw_2015}, we set \(b=1\). Note that equation \eqref{equ_37} is a low redshift extrapolation of the general equation (1) in  \cite{Furlanetto_2005} and is a good approximation for low redshifts, in particular the redshift  when the string signal is most visible, which is around \(z=12\) given the assumptions made in this work. For higher redshifts the amplitude will deviate due to the redshift dependence of $T_S - T_{\gamma}$, but the increase in amplitude is smaller than the increase in the amplitude of the foregrounds. Hence, we can argue that the low redshift result of Eq. (\ref{equ_37}) is a reasonable approximation at the redshifts for which the cosmic string wake signal is strongest. The implications of this approximation are discussed below Figure 18.

Based on the real space matter power spectrum Eq. \eqref{sim2},  we need to determine the induced angular power spectrum for observations involving integration over a frequency band. We use the flat-sky approximation and find \cite{Shaw_2015} 
\begin{align}\label{sim7}
    C_l(z) \, = \, \frac{1}{\pi\chi \chi'}\int_0^\infty dk_\parallel \cos(k_\parallel \Delta \chi) P_{T_b}(k,z),
\end{align}
where \(\chi\) and \(\chi'\) are the comoving distances for the redshifts \(z\) and \(z'\) marking the lower and upper bound of the redshift bin under consideration. \(\Delta \chi\) represents the differences of the comoving distances. Note that in a fully accurate treatment \(P_m(k,z,z')=P(k)D_+(z)D_+(z')\) where \(D_+(z)\) is the growth factor normalised such that \(D_+(0)=1\). However, since the redshift bins which we consider are very thin, we can use \(P_m(k,z)\) as a good approximation.   Note that for the execution of the integral in Eq. \eqref{sim7} we used the decomposition \(k^2=k_\parallel^2+l^2/\chi^2\), where $k_\parallel$ is the component parallel to the line of sight.

\begin{figure}[!tb]
\centering
\includegraphics[scale=0.9]{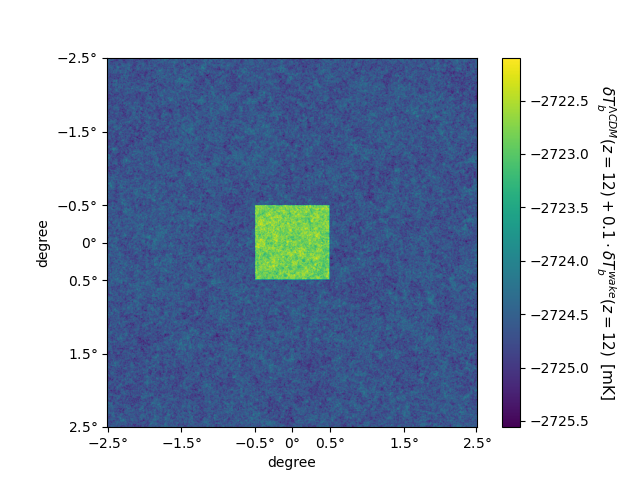}
\caption{Simulation of a box of angular dimensions  $5^\circ \times 5^\circ$ comparing the $\Lambda$CDM fluctuations with the signal of a cosmic string wake (located in the small rectangle in the center of the box). The value of $G \mu$ was taken to be $G\mu = 3 \times 10^{-7}$, and the redshift interval was $ 11.9 < z < 12.1$.}
\label{LCDM}
\end{figure}

In Figure \ref{LCDM} we show the resulting 21-cm brightness map from a realization of the Gaussian $\Lambda$CDM fluctuations, compared with the signal of a string wake with $G \mu = 3 \times 10^{-7}$. Locally in position space the string signal is much larger in amplitude than the contribution from the $\Lambda$CDM noise, the reason being that the string wake is already a nonlinear fluctuation at high redshifts, while the $\Lambda$CDM fluctuations are still in the linear regime. This demonstrates that, provided foregrounds can be taken care of, high redshift observations provide an exciting venue to identify string signals.

Concerning cosmological signals, we consider solely the string wake signature, the four foreground contaminants and the primordial \(\Lambda\)CDM-fluctuations in this work. We assume that all other sources of significant noise disturbance have been eliminated using some sort of subtraction techniques leaving behind the above described smooth residuals. Hence we completed the discussion of the cosmological signals picked up by an interferometer. Let us now turn to the measurement related noise.

\subsection{Interferometer effects}\label{sec_interfer_ch_simulations}

Our aim is to simulate a realistic measurement of a patch in the sky by an interferometer. Such an instrument possesses certain characteristics that have a severe influence on the spectrum of Fourier modes that can be observed as well as on the intrinsic instrumental noise level. In this section, we  analyse the implications of these interferometer effects for our simulations and compare them with the foreground contaminants.

In general, the instrumental noise contains two main contributions. First, the instrument itself, depending on the hardware design, produces a natural noise contribution due to e.g. finite temperature effects or effects due to dust particles. Second, atmospheric effects and background radio sources can further increase the instrumental noise level. However, note that the systematic noise, containing instrumental and sky effects, is typically small compared to the residual foregrounds \cite{Bull_2015}.

The exact specifications of the instrumental noise are quoted in the design specifications for a given experiment. In this work, we adapt a general framework and calculate the instrumental noise power spectrum based on an approximate approach that is easily adaptable to an arbitrary configuration of antennas. Note that this approach holds only for array interferometers (dish or antenna) and not for single dish observations (although the adaptations necessary are minimal). The advantage of this type of surveys is that measurements can be carried out using multiple baselines at the same time and that these are usually much larger than for single dish surveys. Additionally, we focus on antenna array interferometeres like MWA or LOFAR. Since antennas are cheaper and easier to implement, these surveys usually contain a much larger number of baselines and consequently, by the principles of interferometry, can probe a larger domain of Fourier modes. The approach for the instrumental noise that is chosen here, however, remains identical for dish and antenna interferometers.

The instrumental noise is modeled via a power spectrum following \cite{Alonso_2017},
\begin{align}\label{sim1}
    P_T(l) \, = \, \frac{\lambda^2 T^2_{sys} N_p}{A_e^2 \Delta \nu t_{tot} n(u=l/2\pi)}.
\end{align}
Here, \(\lambda\) is the wavelength of the measured radiation, \(T_{sys}=T_{inst}+T_{sky}\) is the system temperature containing the instrumental hardware design dependent temperature \(T_{inst}\) and \(T_{sky}\approx 60 \text{K}\cdot (\nu/300 \text{MHz})^{-2.5}\), the sky temperature accounting for atmospheric effects and background radio emission. Note that usually, the instrumental, or equivalently called the receiver temperature is suppressed compared to the sky temperature \cite{Franzen_2016}. \(N_p\) is the number of pointings and effectively describes the relation between the field of view (FOV) of an interferometer and the fraction of the sky probed by it. \(A_e\) stands for the effective collecting area per antenna or per tile in case the antennas are combined to form antenna tiles (e.g. for MWA). For dish interferometer, it describes the effective total dish area. The total integration time is given by \(t_{tot}\), the bandwidth of an measurement by \(\Delta \nu\). The most fundamental and interferometer dependent part of the power spectrum is the baseline density \(n(u)\) where \(u\) is given by the multipole moment divided by \(2\pi\). It depends on the exact distribution of the antennas with respect to each other and determines for every mode number \(l\) the number of baselines able to measure this mode. Moreover, the baseline density incorporates the natural limitations of an interferometer regarding the minimal and maximal Fourier modes that can be resolved by it. In general, interferometer surveys have upper and lower limits regarding the modes in \(k\)-space\footnote{Note that talking about \(k\)-space in this work refers to, unless explicitly stated otherwise, \(k_\perp\)-modes.} that they can resolve. For large \(k\), an experiment is not able to resolve modes larger than the \(k\)-mode corresponding to the scale of the longest baseline. For small \(k\), only \(k\)-modes bigger than the mode corresponding to the FOV can be measured with a reasonable amount of noise. The modes beyond these limits are strongly contaminated by instrumental noise and thus practically undetectable. This behavior is included in the power spectrum Eq. \eqref{sim1} via the baseline density \(n(u)\) which goes to zero as we approach these modes and hence leads to a divergence of the power spectrum. The baseline density depends on the relative positioning of the antennas or dishes with respect to each other in an interferometer experiment. Thus, to model it accurately, we need a given configuration of antennas as, for instance, from a real experiment. We will come back to this point in section \ref{sec_implement_simulations}.

With the description of the instrumental noise power spectrum we now assembled all signal and noise components relevant for the simulation of a patch in the sky measured by an interferometer. As emphasized before, the foreground noise is expected to be much larger than our targeted cosmic string signal. Hence, we need to define and apply techniques that enhance the string wake signal in the measurement data. Before doing so, let us first introduce the statistics with which we aim to analyse the data. This way we can adapt potential data processing techniques to the statistic to achieve an optimized result for the string detectability.

\subsection{Statistics}\label{sec_statistics_simulations}

Due to the resolution limits in angular and frequency direction, interferometers output their data in discrete pixel space and in redshift bins. The pixel map can be understood as a sample of data points of radiation intensity. We want to analyze this data sample for correlations between specific groups of pixels, i.e. the wake signal. In this work, we focus on two specific statistics. 

\subsubsection{\boldmath\(\chi^2\)-statistics}

A  commonly used statistic to extract a signal from a given background model is the \(\chi^2\)-statistics. Similar to the variance, for every data point the deviation from the mean of the model is calculated and the results are summed up. The resulting estimator contains information about the goodness of the modeled mean which translates to the goodness of the model theory, which consists of an imposed covariance matrix \(C\) and a prediction for the mean \(\mu_i\) for each data point \(i\) (here corresponding to pixels). 

Let \(x_i\) be the observed value of a pixel \(i\) and \(\mu_i\) the corresponding modeled mean value. Further, let us define \(C\) as the model covariance matrix between the pixels such that \(C_{ij}\) represents the correlation between \(x_i\) and \(x_j\), \(C_{ij}=E[(x_i-\mu_i)(x_j-\mu_j)]\). In the latter equation, \(E\) denotes the expected value of its argument. Note that the standard deviation is given by \(\sigma_i=C_{ii}^{\,\,1/2}=E[(x_i-\mu_i)^2]^{1/2}\). Then, the \(\chi^2\)-estimator for a one-dimensional chain of pixels is defined as 
\begin{align}\label{sim3}
    \chi^2 \, = \, \sum_{j=0}^\nu\sum_{i=0}^\nu(x_i-\mu_i)(C^{-1})_{ij}(x_j-\mu_j),
\end{align}
where \(C^{-1}\) is the inverse of the covariance matrix and \(\nu\) represents the degrees of freedom in the data sample. 

In the case of interferometer data, the mean \(\mu_i\) of the background model corresponds to the observational mean of a given type of foreground contaminant and \(C\) to the estimated covariance matrix. Note that this matrix for a given foreground type can be determined by applying the definition above on multiple numerical foreground realizations.\footnote{The concrete application of this statistic is described in section \ref{sec_implement_simulations}.} Equation \eqref{sim3} can further normalize the estimator by dividing by the degrees of freedom so that \(\widetilde{\chi}^2 = \frac{\chi^2}{\nu}\). For testing data with respect to some hypothetical model via the \(\chi^2\)-statistics we need to formulate a model parameter \(\mu\) and the matrix \(C\) in advance. If the chosen model, i.e. \(\mu\) and \(C\), is an accurate approximation of the data that is measured, \(\widetilde{\chi}^2\approx 1\). The \(p\)-value for the acceptance of a hypothetical model based on the measured data can be calculated using the \(\chi^2\)-distribution
\begin{align}
    P(\chi^2) \, = \, \frac{1}{2^{\nu/2}\Gamma(\nu/2)}(\chi^2)^\frac{\nu-2}{2}\exp(-\chi^2/2),
\end{align}
where \(\Gamma\) is the Euler-\(\Gamma\)-function.

The definition above in Eq. \eqref{sim3} holds for real space data. Let us transform this formulation into Fourier space in order to avoid the calculation of the full covariance matrix in real space. In Fourier space, the covariance matrix \(C\) becomes diagonal, and we use the relation between the power spectrum and the Fourier transform \(P(k)\sim |F(k_x,k_y)|^2\) to rewrite Eq. \eqref{sim3} as 
\begin{align}\label{sim4}
    \chi^2 \, \propto \, \sum_i^\nu \frac{{P}_{data}(k_i)}{P_{model}(k_i)\cdot \sigma^2}.
\end{align}
In the latter equation, \(k_i\) corresponds to a discretized \(k\)-mode and \(\sigma\) to the real space variance of the background model. Note that for the application of this formula we have to assume that the above data follows a Gaussian distribution and the power spectrum only depends on the amplitude of the \(k\)-mode. 

In section \ref{sec_implement_simulations} we will discuss in more detail the effects of measuring the data and applying this type of statistic in pixel space. However, we emphasise at this point that in Eq. \eqref{sim4} the string wake signal only constitutes a minimal contribution to the overall power spectrum compared to the foregrounds. Thus, we expect that multiple orders of magnitude may have to be removed from the foregrounds amplitude before the wake's alteration of the overall data power spectrum is sufficiently large for the \(\chi^2\) estimator to show a difference from the foreground model. In this work, we will use the \(\chi^2\)-estimator in Fourier space.   

As a first approach, the \(\chi^2\) provides a good intuition of the signal-to-noise ratio and is comparably simple to implement. Nonetheless, it does not pick out specific characteristics of the wake signature and represents a rather general approach. In the following subsection, we will apply the knowledge acquired in Section \ref{ch_cosmicstring} to customize a suitable statistics for string wake detection.

\subsubsection{Higher order correlation functions}

Non-Gaussianities, such as the signature of a cosmic string wake, are a prominent topic in research in cosmology and are, due to the constant technological progress, more and more explored in the \(21\) cm radiation domain (e.g. for recent works \cite{karagiannis2020probing, Sekiguchi_2019}). As mentioned in \cite{karagiannis2020probing}, a common approach to pick out these type of signatures is the usage of higher point functions. In particular in the case of interferometer data, higher order correlation functions in Fourier space are used. Since we considered the foregrounds in an observed patch to be realized as Gaussian random fields, higher order correlation functions (reduced by their two-point components) without any non-Gaussianity present average to zero when averaged over an large region in space or a large number of samples.  Adding the string wake into the data yields an non-trivial value for the reduced n-point functions in Fourier space.  

The lowest order statistic to which Gaussian noise sources does not contribute is the 3-point function. In order to directly test for the ridge-like features a cosmic string wakes signal exhibits in Fourier space, we choose a particular shape:
\begin{align}\label{sim5}
     \braket{T(\Vec{k}_1)T(\Vec{k}_2)T(\Vec{k}_3)}\,\,\text{with}\,\,\Vec{k}_1\approx-\Vec{k}_2,\,\,|\Vec{k}_1|\approx |\Vec{k}_3|\,\,\text{and}\,\, \Vec{k}_1\cdot \Vec{k}_3\approx 0.
\end{align}
This 3-point function has an additional advantage. We saw that cosmic string wakes can take different shapes in Fourier space which all share the ridge-like arms as a feature. However, these arms can enclose different angles relative to each other, depending on the wakes alignment and on the redshift bin thickness, i.e. how the string wake crosses the redshift bin. The 3-point function can be adapted to these more general alignments simply by choosing \(k\)-vectors that adequately represent the alignment of the ridges with respect to each other. For instance, in the case of the projected string wake forming a parallelogram in real space, the ridges in Fourier enclose the same angle as the edges of the parallelogram. Let this angle be \(\alpha\). Then we have to adapt the conditions for the alignment of the mode vectors in Eq. \eqref{sim5} to \(\Vec{k}_1\cdot\Vec{k}_3\approx |\Vec{k}_1||\Vec{k}_2|\sin{\alpha}\).

\subsection{Signal processing and Foreground Removal}\label{sec_filters_simulations}

A common problem in radio cosmology is the low signal-to-noise ratio of the raw data. Increasing this ratio has become a field of research itself and many image processing techniques and foreground removal strategies (see references in section \ref{sec_fg_ch_simulations}) have been explored. Filtering the data in favor of the string wake signal and removing parts of the residual foregrounds described in section \ref{sec_fg_ch_simulations} are two key steps which we implement. Filtering techniques are common and will be reviewed in the Appendix. The exact functional description of the wake signal in Fourier space as well as the smooth frequency dependence of the residual noise sources can be exploited. Note that the techniques used in this paper are only a few of a vast variety of signal processing tools. Other potentially more powerful and more involved techniques such as the application of machine learning to the data set are used in the literature as well. However, applying these methods is not within the scope of this work.

Foreground removal techniques are a prominent and current research topic, in particular for \(21\) cm maps (see e.g \cite{removal1, Mertens_2018}, and \cite{Liu} for a review). For cosmological studies of the early universe, foreground removal is essential for measuring any cosmological signal as the foregrounds are much brighter than the signals we are after.  In this subsection, we present a suitable foreground removal strategy that we implement in the numerical simulations described later on.

A feature distinguishing between the string wake signal on one hand and galactic as well as extra-galactic foregrounds on the other is their extent in redshift direction. While the foregrounds are smoothly correlated over a large number of redshift bins, the string signal is expected to extend only over a small number of them. In the specific case considered in this work, the string wake signal resides in a single redshift bin. If we analyze an individual pixel of the sky patch measured by an interferometer in multiple consecutive redshift bins, the brightness temperature  displays a smooth redshift dependency, as sketched in Figure \ref{fig:27}. The interpolation of the values over multiple redshift bins will capture the frequency dependence of the foregrounds without being spoiled by the wake signature within the data, as the latter only affects a single redshift bin. This allows for a removal of the foreground contaminants by subtracting the interpolated values, and  will conserve the impact of the wake  on the measured patch. Each angular pixel has to be interpolated individually as the scaling in redshift can vary from pixel to pixel. In the context of interferometer surveys, this method is well applicable as interferometers measure data in a frequency band which includes multiple neighbouring redshift bins at the same time.
\begin{figure}[!tb]
\centering
\includegraphics[scale=1.1]{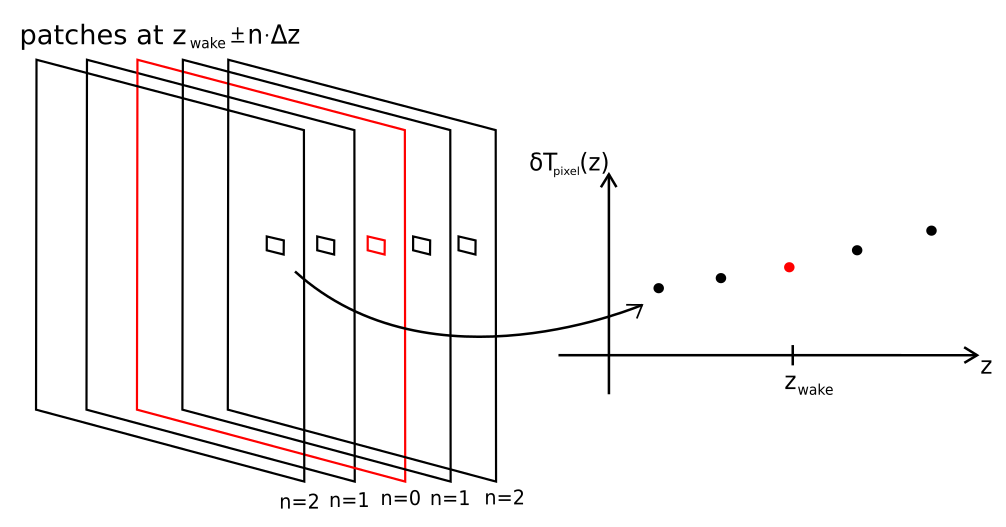}
\caption{Sketch of the foreground removal procedure. On the left hand side five patches corresponding to five consecutive redshift bins are lined up. The red patch contains the string signal. For each patch we mark the same single pixel. The brightness temperature of these pixels is then plotted on the right hand side as a function of redshift. }
\label{fig:27}
\end{figure}

\begin{figure}[!tb]
  \centering
  \begin{minipage}[b]{0.5\textwidth}
    \includegraphics[width=\textwidth]{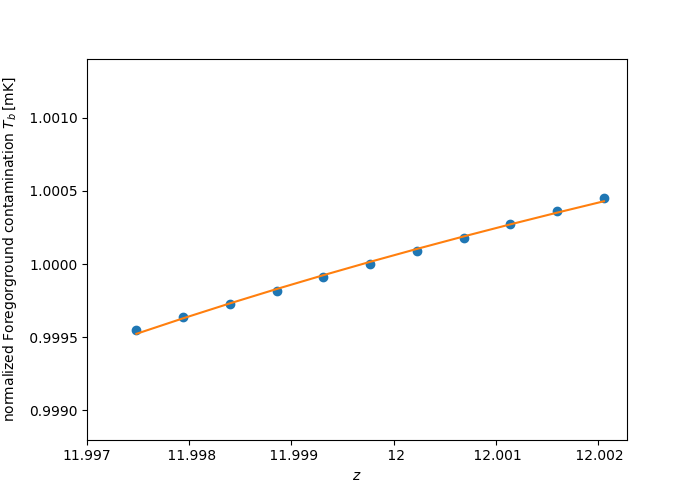}
    \caption{Normalized pixel brightness temperatures for a map containing the four residual foreground components discussed in section \ref{sec_fg_ch_simulations}, \(\Lambda\)CDM and instrumental noise. The pixels are extracted from multiple patches as sketched in Figure \ref{fig:27}. We normalize with respect to the mean of the displayed brightness temperature values. The dots mark the normalized foreground value at the center of the corresponding redshift bins whose width in units of frequency is given by \(50\) kHz. The yellow line represents the fit function \(\sim (1+z)^\alpha\).}
    \label{fig:22}
  \end{minipage}
  \hfill
  \begin{minipage}[b]{0.49\textwidth}
    \includegraphics[width=\textwidth]{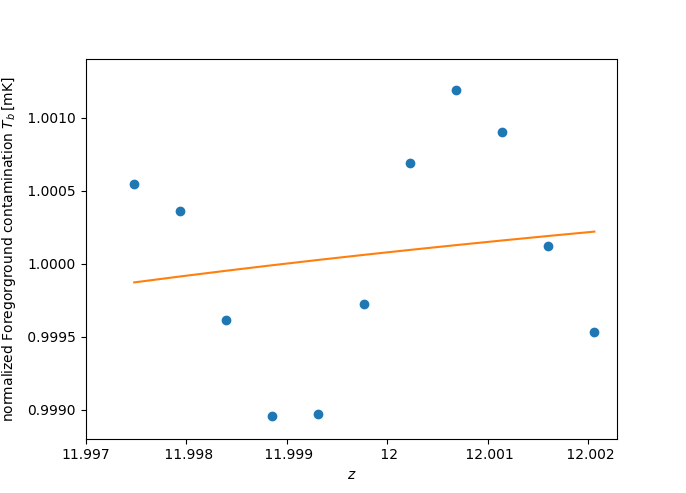}
    \caption{Normalized pixel brightness temperatures as in Figure \ref{fig:22}. On top of the foregrounds we add instrumental effects in the form of noise with a sine-shaped frequency dependence. The yellow line represents the fit function \(\sim (1+z)^\alpha\).\\\\\\\\\\}
    \label{fig:23}
  \end{minipage}
\end{figure}

The interpolation of the residual foregrounds in the considered patch is possible due to the foregrounds being smooth in frequency direction. In fact, following the discussion in subsection \ref{sec_fg_ch_simulations}, they display a frequency dependence with a fixed exponent. We can use this behavior and construct multiple consecutive redshifts bins following the functional redshift dependency given in Eq. \eqref{porcozio}. Additionally, we add uncertainties in the exponent according to observational data (as in \cite{numerics7}). We simulate \(11\) frequency bins, and insert the signal into the frequency bin in the middle. Further, we add \(\Lambda\)CDM-fluctuations and instrumental noise on top.\footnote{The exact implementation will be discussed in section \ref{sec_implement_simulations}.} For each pixel in these \(11\) maps we line up their brightness temperature values, and fit a function of redshift to these values, as displayed in Figure \ref{fig:22}. This function takes a certain value for the redshift bin which includes the signal. For this bin, we can subtract the interpolated foreground vales and repeat the procedure for the next pixel in the patch. As a result, we can approximately remove the foreground from the patch including the string wake.

In general, this method removes the foregrounds very accurately. However, there is one caveat: Instrumental effects of interferometer surveys may respond to a sky signal which has oscillations in frequency direction. The oscillation amplitude scales with the amplitude of the foreground. This effect may induce a mixing of the high amplitude foregrounds with the low amplitude signal which we aim to measure. Even if the mixing is small, the response function may have a significant impact on the detectability of the signal due to the much brighter foregrounds. Since the response is generally unknown, it requires a high degree of instrument calibration and thus poses a major challenge for interferometer surveys. Here, we simulate this behavior with a sine-shaped noise with an amplitude of approximately one percent of the foregrounds amplitude. Testing different periodicities for this sine-noise, we cover a group of potential response functions of interferometers and can give quantitative feedback about the detectability of the string wake for each of them. In Figure \ref{fig:23} we displayed a sine-shaped noise on top of the foregrounds for each redshift bin. Depending on the periodicity, the interpolation of the foreground may be less accurate. Note that in order to better compare the pixel interpolation, we normalized the pixel brightness temperatures of Figure \ref{fig:22} and \ref{fig:23} by the mean of the values without the sine-shaped noise. When comparing the two figures we see a slightly different tilt in the fit function due to the instrumental response function in Figure \ref{fig:23}.

In the following we turn to a description of our implementation of these techniques in the simulation program. 

\subsection{Implementation}\label{sec_implement_simulations}

\subsubsection{Methodology}

The numerical simulations of this work are implemented in the open-source programming language {\fontfamily{lmtt}\selectfont Python}.  Most of the preexisting functions and methods we apply in this analysis are standard functions in {\fontfamily{lmtt}\selectfont Python} and do not require further introduction. In particular, we have used the {\fontfamily{lmtt}\selectfont PyCosmo}-package \cite{PyCosmo} that was developed at the Institute for Particle Physics and Astrophysics of the ETH Zurich. 
\hyperlink{https://github.com/maibachd/simulationthesis}{\text{GitHub}} \cite{Git}. 
Some of the calculations were executed using the ETH computer cluster  \hyperlink{https://scicomp.ethz.ch/wiki/Euler}{{\fontfamily{lmtt}\selectfont Euler}}.  

\subsubsection{Signal and noise components}

In our analysis we will focus on Fourier maps obtained from position space maps in one redshift bin. Given a three-dimensional map, it means that for each pixel in angular space we integrate the signal in redshift direction over the bin.

Our analysis assumes that the string wake resides in a single redshift bin. Consequently, in real space the projected shape corresponds to a rectangle. Note that due to the increasing thickness of the wake in real space moving away from its tip, the brightness temperature signal obtains a gradient in intensity across the rectangle. For simplicity, we neglect this effect and assume that the string wake is square-shaped and of homogeneous intensity., the intensity being obtained by integrating the brightness over the redshift extent of the wake's image. The size of the square in angular directions is fixed to be \(1^\circ \times 1^\circ\) in the sky. This corresponds to the angular size expected from a Hubble length string segment moving relativistically and which was laid down at \(t_{rec}\).  We include the string signal of the above size in a patch of \(5^\circ \times5^\circ\). 

Given this configuration in real space, we obtain the Fourier transform map by dividing the real space map into \(512 \times 512\) pixels and performing a Fast Fourier transformation (making use of the extension package {\fontfamily{lmtt}\selectfont Numpy}). This analysis corresponds to an angular resolution of \(36''\), a resolution which can be reached by modern day interferometers such as the SKA \cite{SKA}.  

The foreground contamination per pixel corresponds to \(\Delta z \cdot T_b^{fg}\) where \(\Delta z\) is the thickness of the redshift bin and \(T_b^{fg}\) is the brightness temperature of the residual foregrounds. In order to implement the residual foregrounds according to their power spectra we calculate the value for each pixel in Fourier pixel space as the sum over the individual foreground components. Each contaminant is individually implemented as
\begin{align}\label{sim8}
    T_b^{fg_i}(k_x,k_y) \, = \, \sqrt{\frac{1}{2}\cdot C_l^i(k_x,k_y)}\bigg(g_1(k_x,k_y)+i\cdot g_2(k_x,k_y)\bigg),
\end{align}
where \(g_1\) and \(g_2\) are Gaussian random coefficients (normal distributed: \(\mu=0\), \(\sigma=1\)) and \(C_l^i\) corresponds to the angular power spectrum for the residual foreground component \(i\) (defined in Eq. \eqref{porcozio}). As we work in the flat sky approximation, the \(l\) in the angular foreground power spectrum given by Eq. \eqref{porcozio} is replaced with \(l=(k_x^2+k_y^2)^{1/2}\). Subsequently, we normalize each contaminant according to their observational value \cite{numerics7} by setting the value of \(k=0\)-mode in Fourier pixel space equal to the real space mean and adjusting the real space standard deviation using Parseval's theorem. In the case of discrete space, this theorem reads 
\begin{align}
    \sum_{n=0}^{N-1}|x[n]|^2 \, = \, \frac{1}{N}\sum_{k=0}^{N-1}|X[k]|^2,
\end{align}
where \(X[k]\) is the discrete Fourier transform of \(x[n]\), both of length \(N\). Note that for the two-dimensional case, slight adaptations are necessary. After this normalization procedure for every contaminant and adding together all foreground component brightness temperature maps, we obtain a real space patch with the residual foreground contamination in agreement with observational data. 

For the galactic foregrounds, i.e. Synchrotron and free-free emission, we display the real space patches of the residual foreground contamination in Figures \ref{fig:25} and \ref{fig:26}.
\begin{figure}[!tbp]
  \centering
  \begin{minipage}[b]{0.49\textwidth}
    \includegraphics[width=\textwidth]{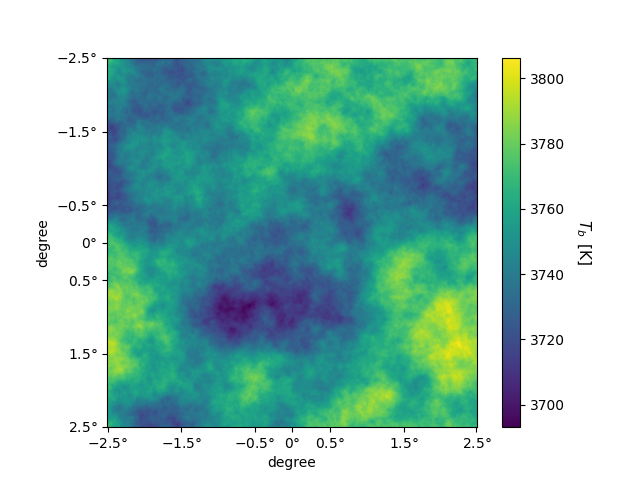}
    \caption{Simulated real space map of the total brightness temperature with \(512 \times 512\) pixels and angular resolution of \(36''\). The patch is filled with a Gaussian realization of galactic Synchrotron radiation at a redshift of \(z=30\).}
    \label{fig:25}
  \end{minipage}
  \hfill
  \begin{minipage}[b]{0.49\textwidth}
    \includegraphics[width=\textwidth]{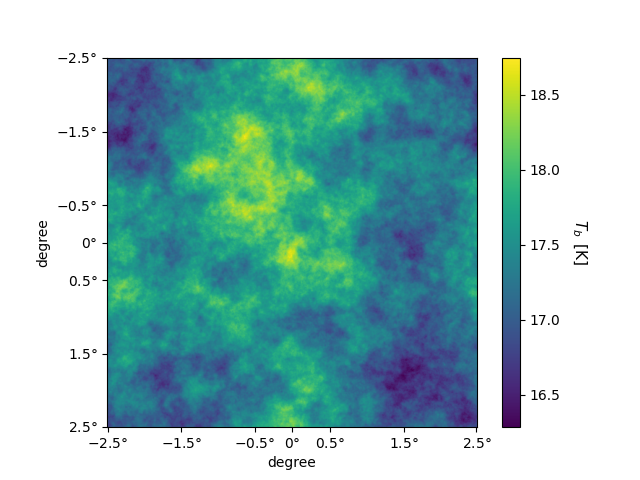}
    \caption{Simulated real space map of the total brightness temperature with \(512 \times 512\) pixels and angular resolution of \(36''\). The patch is filled with a Gaussian realization of galactic free-free radiation at a redshift of \(z=30\).}
    \label{fig:26}
  \end{minipage}
\end{figure}

Additionally, we want to include \(\Lambda\)CDM-fluctuations in the patch. They are also described by a Gaussian random field, and, making use of Eq. \eqref{sim7}, we can construct the pixel values in analogy with (\ref{sim8}). Note that the primordial matter power spectrum in Eq. \eqref{sim7} is calculated using the {\fontfamily{lmtt}\selectfont PyCosmo}-package. We use the value of the \(\Lambda\)CDM-fluctuation at the redshift of the bin being considered. Like in the computation of the wake signal, we determine the relative brightness temperature of primordial fluctuations with respect to the CMB background radiation. The relative brightness temperature is calculated via \(\delta T_b(z)=(T_b(z)-T_\gamma(z))/(1+z)\) \cite{Furlanetto_2006} and we find that the \(\Lambda\)CDM signal is much weaker at redshift \(z=12\) compared to the foregrounds. The brightness temperature fluctuations in the Gaussian realization are \(\sim\mathcal{O}(10)\) mK. This yields an average relative brightness temperature \(\delta T_b\) of approximately \(2.724\) K in absorption. This amplitude is larger than the extragalactic free-free emission (\(\mu_{T_b}(z=12)\approx 280\) mK) but smaller than all other contaminants. Note that without foregrounds, the string wake can easily be detected against \(\Lambda\)CDM-fluctuations at high redshift, as shown in Figure \ref{LCDM}.

\begin{figure}[!tb]
\centering
\includegraphics[scale=1.1]{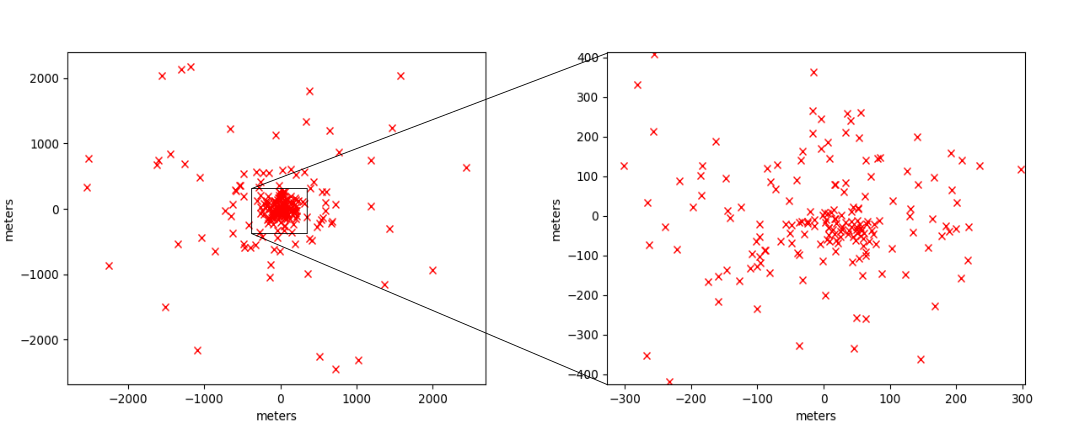}
\caption{Relative alignment of MWA's \cite{MWAI, MWAII} Phase I and II antenna tiles. On the right side we zoomed into the most densely populated area of the interferometer configuration.}
\label{fig:24}
\end{figure}

Finally, we implement the instrumental noise. This noise is highly dependent on the type of interferometer and the survey mode. Technically,  Eq. \eqref{sim1} is arbitrarily adaptable, however, for the purpose of this work we use the MWA configuration as a benchmark. The MWA survey appears suitable for string detection because the instrument is in a radio quiet location, it has good frequency resolution, and because of the large number of baselines. We assume a frequency bandwidth of \(50\) kHz and an integration time of \(1000\) hours. The system temperature is defined below  Eq. \eqref{sim1}. Phase I and II of MWA together consist of \(256\) antenna tiles each consisting of \(16\) antennas. The effective collecting area per tile is given by \(A_e\approx21.5\) square meters. In order to calculate the baseline density \(n(u)\) we need an exact distribution of antennas and their positions relative to each other.  We were provided with this data by Cathryn Trott from the MWA Epoch-of-Reionization team. The relative alignment of the MWA antenna tiles is displayed in Figure \ref{fig:24}. Note that smaller baselines are much more numerous than long ones. Consequently, the noise grows significantly at an intermediate mode and diverges when too few baselines are available. We numerically model divergent instrumental noise by dropping the pixels in which the noise exceeds a certain large threshold. 

Based on the data provided by Cathryn Trott, the baseline density can be calculated. As this is a rather complicated procedure, we adapted a pre-existing program created by one of us (DC). Based on the input of the relative locations of the antennas or dishes, the program outputs the baseline density \(n(u)\) provided we adapt a few variables. The result is influenced by the numerical parameters which describe the fit function used to interpolate the baseline densities at different scales. We can model the fit function in two ways. Either we allow for a low noise level but fewer resolved modes or we accept a large noise power spectrum resolving more modes. Within the region of non-divergent noise power spectrum, the interferometer noise is much smaller than the foregrounds and the detectability is more significantly affected by the restriction in the resolved modes. We hence chose a pessimistic fit function with respect to modes that are not affected by the divergence of the noise, i.e. we chose a fit function for which the interferometer resolves fewer modes with finite noise. As one of our goals is to investigate interferometer configuration favoring the string wake detection, it is reasonable to start out with strict constraints coming from the interferometer noise power spectrum. Based on these, we can analyze detection improving parameters. In general, the instrumental noise is simulated as being Gaussian in Fourier space but it is negligible at small scales compared to the foregrounds. Thus, the strongest restriction induced by instrumental noise originates from the divergence of the noise at certain \(k\)-scales, as explained in section \ref{sec_fg_ch_simulations}. Note that around the \(k=0\)-pixel we additionally cut away the smallest \(k\)-modes as they can generally not be measured robustly due to instrumental effects either. In the flat sky approximation, i.e. using \(k/180^\circ=l/\pi\), this low-\(k\)-cutoff corresponds to ignoring contributions of angular modes \(l<200\) in the calculation of the statistical estimators.

\subsubsection{Statistic, filters and removal techniques}

After the implementation steps listed above, we obtain a patch in Fourier space containing four residual foreground contaminants, the string wake, \(\Lambda\)CDM-fluctuation and instrumental noise. Note that this would not correspond to the raw data stream coming from an interferometer since we assume that some kind of foreground removal technique has already been applied (as discussed in section \ref{sec_fg_ch_simulations}). This removes e.g. extremely bright point sources, leaving only the residual foreground components described in earlier sections in the data. Next, the patch is subject to the filters and foreground removal procedures described in section \ref{sec_filters_simulations}. 

The filters are implemented as described in section \ref{sec_filters_simulations}. Given the power spectra of all of the noise sources (including the $\Lambda$CDM ``noise''), which are all modelled as Gaussian random processes, we set up a realization of these noise sources in our Fourier space patch.   

The implementation of the removal of the foregrounds was described in Section \ref{sec_filters_simulations}. In the simulations, we consider \(11\) distinct redshift bins surrounding the one which is being analyzed. We fit the redshift dependence of the signal to a form \((1+z)^\alpha\) where \(\alpha\) serves as a fit parameter capturing the smooth redshift dependence of the sum of the foreground contributions listed in Table 1 in the simulated patch. Note that, as aforementioned, when generating the foregrounds we include uncertainties in the form of choosing the redshift scaling of the individual foreground components for each redshift bin and pixel to be Gaussian normal distributed \(\mathcal{N}(\mu_i, \sigma)\), where the mean \(\mu_i\) is the redshift scaling of the contaminant \(i\) given by the corresponding value in Table \ref{tab:1} and \(\sigma =0.1 \) for all foreground types. We adapted the value of \(\sigma\) from the standard deviation of the redshift exponent of galactic Synchrotron radiation in \cite{numerics7}. The contamination from residual foregrounds and primordial fluctuations is interpolated for the redshift bin in the middle of the \(11\) slices of redshift space, i.e. for the one containing the string wake. This interpolation value is subsequently subtracted from the pixel value. The procedure is repeated for each pixel.

With the removal of the foregrounds and the application of the filters\footnote{We first remove the foregrounds and then apply the filters in this work.} the signal to noise ratio in the analyzed data patch is significantly increased. At this point the statistics to search for the string wakes signal can be applied. A good statistic should be able to reveal the presence of the string wake at a high statistical significance. However, we can also artificially shrink the amplitude of the residual foregrounds by choosing a foreground removal factor \(\epsilon_{fg}<1\) and study what value of $\epsilon_{fg}$ is needed in order that the string signal becomes visible (for a good statistic the value of $\epsilon_{fg}$ should be one. The \(\chi^2\) statistic and the three-point function with the particular shape described in the previous section are used. Note that for the calculation of \(\chi^2\) following Eq. \eqref{sim4} we bin the patch into \(300\) \(k\)-bins and compare histograms of the model and the data power spectrum for this binning. This results in \(300\) degrees of freedom by which we can normalize the \(\chi^2\) estimator to be approximately one.

We perform the simulation pipeline multiple times and average over the results. The size of the samples is explicitly stated in the section on our results. This sampling, however, requires a long chain of processes to be executed, and as {\fontfamily{lmtt}\selectfont Python} is per default structured in the way that each process is carried out successively this is a time intensive task. Therefore, we apply a multi-processing routine in which the processes are explicitly separated and executed in parallel, i.e. realizations of the patch are generated in parallel and joined at the end before averaging. This results in simulations of reasonable sample size and simultaneously adequate computational timescales.

\subsubsection{Optimal detectability}\label{subsec_optdect_simulataions}

The redshift dependence of the string wake signal and of the foreground noise sources are very different. As derived in Eq. (\ref{SW12}), the brightness temperature of the wake signal scales as $(1 + z)^{1/2}$ up to the time when the $\Lambda$CDM nonlinearities disrupt the wake. On the other hand, the foregrounds increase in amplitude faster as the redshift increases. Thus, we would expect the string wake signal to be most easily detectable at the lowest redshift before wake fragmentation. A complicating factor is that the string wake signal shifts from absorption at high redshifts (when the kinetic temperature of the wake is lower than the temperature of the CMB) to emission at lower redshifts. Here, we investigate for which redshift the signal to noise ratio is maximized. This will determine the choice of redshift for our actual analysis in the following section.

The total redshift domain considered here is \(z\in [12,30]\). The lower bound is set by the onset of the epoch of reionization, the upper bound by the limitations of most of the currently operating interferometer instruments. 

The signal to noise ratio in interferometer data can be expressed by the intensity ratio
\begin{align}\label{sim9}
    D_{wake} \, = \, \frac{|\delta_z T_b^{wake}|}{|\Delta_z T_b^{fg}|},
\end{align}
where we multiplied the temperatures with the thickness in redshift space as we integrate over the redshift bin. For the brightness temperature of the foregrounds (derived from their power spectrum in Eq. \eqref{porcozio}) we use a particular mode number \(l\) whose choice does not affect the redshift dependence of the ratio Eq. \eqref{sim9}.  

For the value $G\mu = 3 \times 10^{-7}$, the normalized fraction (\ref{sim9}) is plotted in Figure \ref{fig:25} for various values of the redshift $z_i$ of wake creation. The curves $T_b^{diff}$ are obtained taking into account the fact that a wake changes for shock-heated to diffuse at higher redshifts, the curves $T_b$ assume shock-heating for all redshifts.
\begin{figure}[!tb]
\centering
\includegraphics[scale=0.3]{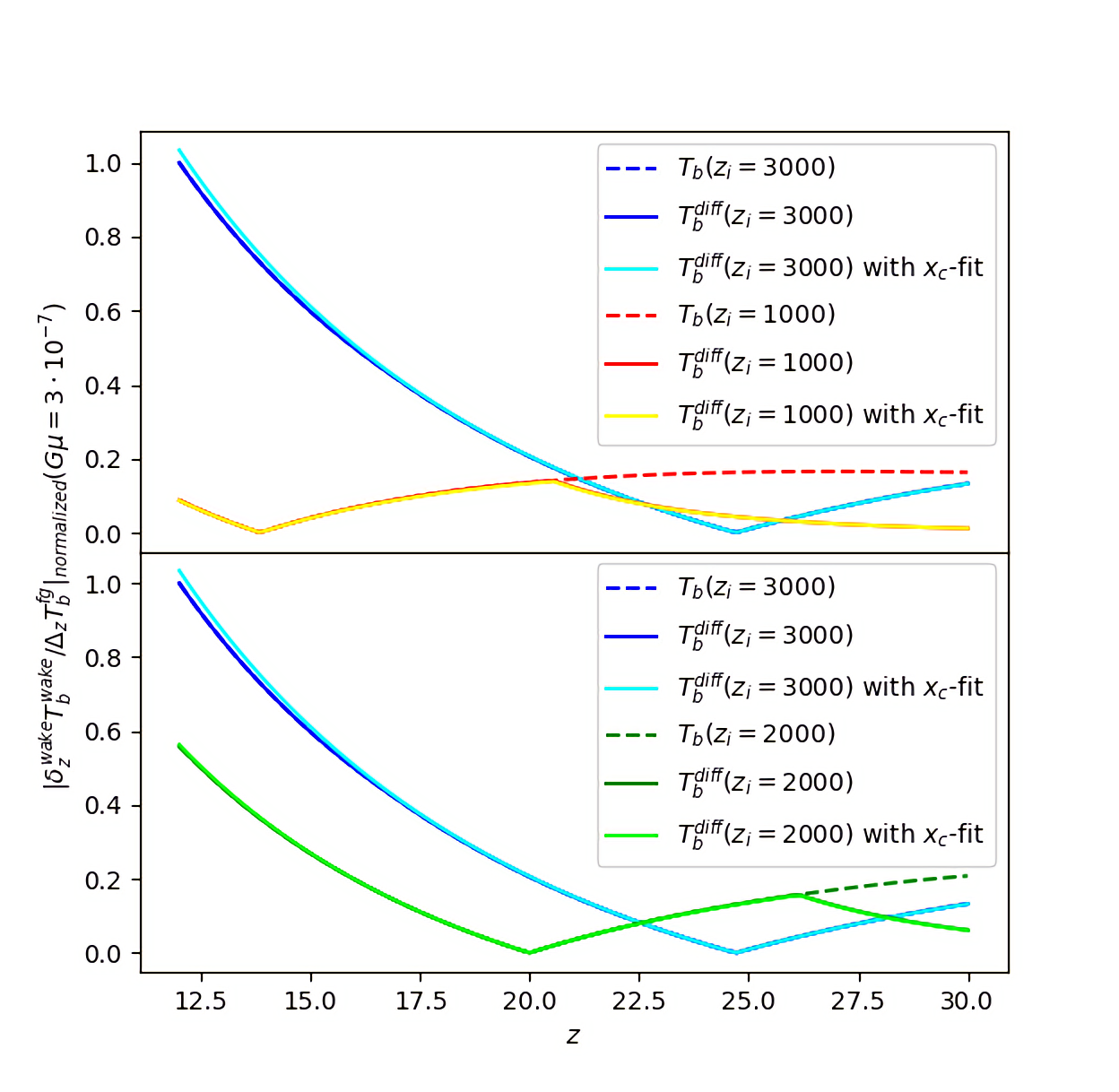}
\caption{Normalized signal to noise ratio following  Eq. \eqref{sim9} for \(G\mu = 3 \cdot 10^{-7}\), \((v_s\gamma_s)^2=1/3\), \(\theta=0.32 \cdot \pi\) and for various values of the redshift \(z_i\) of wake generation. The curves \(T_b^{diff}\) take into account the fact that the wakes become diffuse at higher redshifts, the \(T_b\) curves show the results if diffusion is ignored and shock heating is assumed to occur for all redshifts. We additionally plot the temperature curves for a linear interpolation of collision coefficient \(x_c\) to show the impact of the interpolation proposed in section \ref{sec_theo_cosmicstring}.}
\label{fig:25}
\end{figure}
Here, we assumed the angle \(\theta\) of the string wake to be optimal for string detectability. In this idealized scenario the string wake leads to a square-shaped homogeneous block of extra emission or absorption positioned in a single redshift bin.  In Figure \ref{fig:25} the absolute value of the signal to noise ratio is displayed. Note that the redshift where the ratio (\ref{sim9}) vanishes corresponds to the transition point from emission to absorption for the string wake. 

Another kink of the curves in Figure (\ref{fig:25}) occurs at the transition point from shock heating to diffusion dominance of the wake. The redshift of this transition \(z_{trans}\) is determined by
\be
    T_K \, = \, 3T_g \, ,
\ee
where $T_g$ is the temperature of the background gas. This leads to
\be
1+z_{trans} \, = \, \bigg(\frac{20}{0.06} (G\mu\cdot 10^6)^2 \gamma_s^2 v_s^2 (1+z_i)\bigg)^\frac{1}{3}.
\ee
For a string tension \(G\mu= 3 \cdot 10^{-7}\) and a string velocity of \((v_s\gamma_s)^2=1/3\), we find the maximal detectability at redshift \(z=12\), i.e. at the lower bound of the redshift domain, for wakes laid down at approximately \(z_i=3000\). This configuration is used as a simulation benchmark for the numerical analysis of the following section.

For lower values of the string tension \(\mu\) the curves of Figure \ref{fig:25} are shifted to the left and the amplitude decreases.   Consequently, \(z=12\) may not not optimal for all tensions \(\mu\). If shifted sufficiently to the left, the redshift of transition between shock heating and diffusion becomes the new maximum of the signal to noise ratio. As is displayed by the yellow curve in Figure \ref{fig:25}, this new peak occurs for $G\mu = 1 \times 10^{-7}$. We call the tension at which the maximum of Eq. \eqref{sim9} shifts \(\mu_{crit}\).  For the benchmark wake with \((v_s\gamma_s)^2=1/3\) and \(z_i=3000\) we find that the optimal redshift for detectability of the wake signal shifts away from $z = 12$ once the string tension falls below \(G\mu_{crit} \approx 1.8 \cdot 10^{-7}\). Note that the critical tension decreases monotonically in \(v_s^2\gamma_s^2\) and \(z_i\).  For \(G\mu < 1.8 \cdot 10^{-7}\), the maximum of the signal to noise ration occurs at \(z_{opt}=z_{trans}\). Note also that as explained before equation \eqref{equ_37} is a good approximation of the background brightness temperature if \(z=12\) roughly marks the onset of reionization. For higher redshifts \( \bar T_b\) evolves according to Table 1 in \cite{Furlanetto_2005} and reaches \(\mathcal{O}(10)\) mK. Compared to the foregrounds this is a negligible effect so that the signal to noise ratio evolution in redshift is still well described by Figure 18.

The dependence of the optimal signal to noise ratio on the string tension is shown in Figure \ref{fig:26}. The blue curve gives the signal to noise ratio using the redshift bin at $z = 12$, while the red curve takes into account that the optimal redshift for string detectability will change as the string tension is lowered. The red curve has two kinks, the first resulting from the transition from \(z_{opt}=12\) to \(z_{opt}=z_{trans}\). The second kink marks the transition to diffuse wakes.  
\begin{figure}[!tb]
\centering
\includegraphics[scale=0.7]{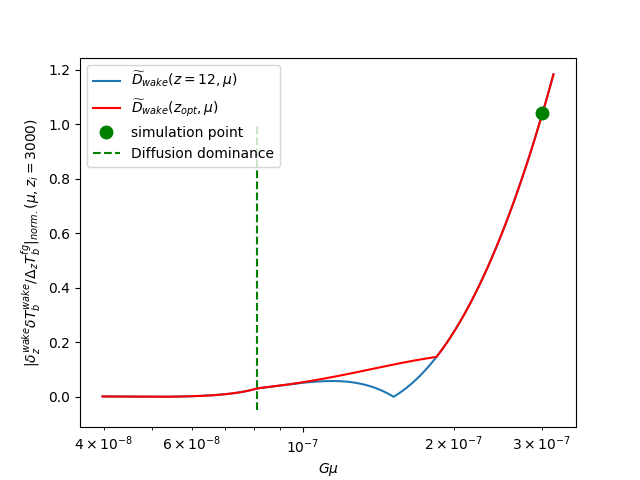}
\caption{Signal to noise ratio Eq. \eqref{sim9} depending on the string tension \(G\mu\). We plot two versions of the ratio, one in which we fix the redshift at \(z=12\) (blue) and one where for every \(G\mu\) the ratio Eq. \eqref{sim9} is evaluated at the optimal redshift \(z_{opt}\) (red), where it is maximal. The green dot marks the point of our simulation, the dashed line marks the tension for which \(z_{trans}=12\). On the left side of the line, diffusion dominates the wake thermalization.}
\label{fig:26}
\end{figure}

\section{Results}
 
 In this section we will study the effectiveness of different combinations of statistics (as described in section \ref{sec_statistics_simulations}) and filtering techniques (see section \ref{sec_filters_simulations}) in identifying the signal of a cosmic string wake in numerically simulated interferometer data.
 
\subsection{Comparing statistics}

As a first step, one wants to identify which of the previously described statistics in general yields a superior string wake detectability. Since foreground removal methods work analogously for both statistics, we can compare the \(\chi^2\)-estimator and the three point function before any application of removal techniques. Note that the instrumental noise is also ignored at first. This noise primarily restricts the number of pixels with convergent noise in the measurement patch. Both statistics in some sense average over the pixel values of the patch and hence we expect them to be equally affected by a restriction of the pixel space due to divergent noise components.

The results discussed now are based on simulations which include all of the effects we have discussed except for instrumental noise. As justified in previous sections, we choose the optimal redshift bin for string detectability. We thus use \(z=z_{opt}(z_i,\mu,v_s)=12\), \(z_i=3000\), \((v_s\gamma_s)^2=1/3\), \( G\mu= 3\cdot 10^{-7}\) and \(\theta\approx 0.32 \pi\) .  

 In Table \ref{tab:2} we show the results of the application of the normalized \(\chi^2\)-statistics for multiple redshifts and angular resolutions. Note that without inclusion of instrumental noise the angular resolution refers to the area-to-pixel ratio used for the simulations. As no specific antenna configuration is incorporated at this stage, the resolution can be chosen arbitrarily. The wake is implemented according to the aforementioned simulation benchmark, the foregrounds (including \(\Lambda\)CDM-fluctuations) are realized following the descriptions in sections \ref{sec_fg_ch_simulations} and \ref{sec_implement_simulations}. For each simulation we list the value of \(\Delta \chi^2=|1-\Tilde{\chi}^2|\) quantifying the deviation of the simulated patch (foregrounds, \(\Lambda\)CDM-fluctuations and string wake signature) from the theoretical background model (foregrounds and \(\Lambda\)CDM-fluctuations). Additionally, we mark the application of a filter \footnote{We used the Wiener filter. Results for matched filtering would be similar.} with a "+" in the "Filter"-column ("-" hence corresponds to no filter application) and calculate a confidence level on \(\Delta\chi^2\) in the form of the \(p\)-value. The foreground removal factor as defined in section \ref{sec_fg_ch_simulations} quantifies the detectability of the wake. 
\begin{table}[t!]
    \centering
\begin{tabular}{ccccccc} \toprule
   row& {angular resolution} &redshift \(z\)& filter & removal factor {$\varepsilon_{fg}$}& \(\Delta\chi^2\) &\(p\)-value \\ \midrule
   1& \(36''\)   &    \(12\) &-    & \(10^{-3}\) & \(0.1281\) & \(0.0712\)\\
   2& \(36''\)   &    \(13\) &-    & \(10^{-3}\) & \(0.0966\) & \(0.1200\)\\
   3&    \(36''\)   &    \(12\) &+    & \(10^{-3}\) & \(0.6689\) & \(<0.0001\)\\
    
 4&    \(1.2'\)   &    \(12\) &-    & \(10^{-3}\) & \(0.1056\) & \(0.1054\)\\
 5&  \(1.2'\)   &    \(12\) &+    & \(10^{-3}\) & \(0.5210\) & \(<0.0001\)\\
 6&    \(36''\)   &    \(25\) &- & \(10^{-3}\) & \(2.3\cdot 10^{-5}\) & \(0.4889\)\\ \bottomrule
\end{tabular}
    \caption{Results of the \(\chi^2\)-estimator applied to a \(5^\circ\times 5^\circ\) patch in the sky containing residual foreground contaminants and \(\Lambda\)CDM-fluctuations as described in section \ref{sec_fg_ch_simulations}, as well as the string wake with a redshift of \(21\) cm absorption \(z=z_{opt}(z_i,\mu,v_s)=12\), redshift of generation \(z_i=3000\), velocity \((v_s\gamma_s)^2=1/3\), tension \(G\mu= 3\cdot 10^{-7}\) and an angle enclosed between the angular plane on the sky and the wake \(\theta\approx 0.32 \pi\). The "+" in the column "Filter" indicates the application of the Wiener-filter before applying the statistic. The \(\Delta \chi^2\) are calculated based on the averages \(\Tilde{\chi}^2\) for a patch with and without string wake for \(100\) random realizations of the foregrounds.}
    \label{tab:2}
\end{table}

Let us now discuss the results. We see from the first row that if the foreground amplitude is suppressed by three orders of magnitude, then, for a detectability benchmark of \(p=0.10\), the string wake is detectable in a patch of \(5^\circ\times 5^\circ\) in the sky at redshift \(z=12\) given an angular resolution of \(36\) arc-seconds. As seen from the second row, the detectability decreases if the redshift increases. From the third row we see that the detectability greatly increases if the Wiener filter is applied. As seen in the 4th and 5th row, the detectability decreases when the angular resolution is worsened. Finally, at \(z=25\) the detectability is almost two orders of magnitude smaller (6th row).

 Based on the results of Table \ref{tab:2}, we conclude that (even without the inclusion of instrumental noise) without application of foreground removal methods, the \(\chi^2\)-estimator applied on realistic interferometer data with an extent of \(5^\circ\times 5^\circ\) in the sky is able to detect the string signature only when the amplitude of the foregrounds is reduced by approximately three orders of magnitude, with or without filter techniques.  

\begin{table}[t!]
    \centering
\begin{tabular}{ccccccc} \toprule
   row& {angular resolution} & redshift \(z\)& filter & removal factor {$\varepsilon_{fg}$}& detect. \(3\sigma\) & detect. \(5\sigma\) \\ \midrule
   
   1& \(36''\)   &    \(12\) &-    & \(10^{-3}\) & + & +\\
   2& \(36''\)   &    \(12\) &+    & \(10^{-1}\) & + & +\\
   3& \(36''\)   &    \(12\) &+    & \(2.5\cdot10^{-1}\) & + & -\\
   4& \(36''\)   &    \(12\) &+    & \(5\cdot 10^{-1}\) & - & -\\
   5& \(36''\)   &    \(25\) &+    & \(10^{-3}\) & + & +\\
    6& \(1.2'\)   &    \(12\) &+    & \(10^{-1}\) & + &-\\\bottomrule
\end{tabular}
    \caption{Results of the three-point function \(\braket{T_b(k_1)T_b(k_2)T_b(k_3)}\)-estimator applied to a \(5^\circ\times 5^\circ\) patch in the sky containing residual foreground contaminants and \(\Lambda\)CDM-fluctuations as described in section \ref{sec_fg_ch_simulations}, as well as the string wake with a redshift of \(21\) cm absorption \(z=z_{opt}(z_i,\mu,v_s)=12\), redshift of generation \(z_i=3000\), velocity \((v_s\gamma_s)^2=1/3\), tension \(G\mu= 3\cdot 10^{-7}\) and an angle enclosed between the angular plane on the sky and the wake \(\theta\approx 0.32 \pi\). The "+" in the column "Filter" indicates the application of the Wiener-filter before applying the statistic. The "+" and "-" in the last two columns indicate weather the three-point function indicates a detectability with a \(3\sigma\) and a \(5\sigma\) confidence level for the given configuration. Note that \(\sigma\) here is defined with respect to a larger survey of \(20000\) square degrees, hence \(\sigma=\sigma_{three-point}/\sqrt{20000/(5\cdot5 )}\). For the listed results, we average over \(10000\) random realizations of the noise contaminants.}
    \label{tab:3}
\end{table}

The three-point statistic (for the special shape chosen to identify the ridges which cosmic string wakes produce in Fourier space) yields significantly better results. The results are listed in Table \ref{tab:3}. Here, we list the confidence level via calculating the difference in the numerical values for the three-point functions with and without the string wake in units of the standard deviation \(\sigma\) of the averaged values. Note here that the \(\sigma\) of the three-point function is based on a large scale survey with \(20000\) square degrees in area. Assuming the area can be divided in uncorrelated patches of  \(5^\circ\times 5^\circ\) in the sky, \(\sigma\) scales as 
\begin{align}\label{res1}
    \sigma \, = \, \frac{\sigma_{three-point}}{\sqrt{20000/25}},
\end{align}
where \(\sigma_{three-point}\) is the actual standard deviation of the averaged three-point function of a single \(25\) square degree patch in the sky. The assumption that these patches are uncorrelated is non-trivial. In real world measurements, the correlation of these patches can potentially yield a bias that has to be modeled separately in the numerical simulation pipeline. For this work however, we assume that these correlations occur only on the largest angular scales corresponding to the smallest \(k\)-scales in the patch in Fourier space. These are cut out in the signal processing pipeline before the calculation of the three-point function, as described in section \ref{sec_implement_simulations}. Thus, for the purpose of this work, we ignore any correlations between patches and the standard deviation can be calculated using  Eq. \eqref{res1}.

The results of Table \ref{tab:3} indicate that, similar to the \(\chi^2\)-statistic, for an angular resolution of \(36\) arc-seconds at redshift \(z=12\), the string wake is detectable to a \(5\sigma\) confidence level if three orders of magnitude are removed from the foreground amplitude. However, applying the Wiener-filter which does not result in a significant enhancement of the detectability using the \(\chi^2\)-statistic yields a two orders of magnitude improvement for the three-point function, as the 2nd row indicates. This constitutes a significant advantage of the three-point statistic over the \(\chi^2\)-estimator.
Further, changing the redshift to \(z=25\) yields a two order of magnitude difference in the detectability for the filtered, as indicated in the 5th row..   

The greater effectiveness of the filters when using the three-point function is not the only advantage of this statistic. Due to the choice of the \(k\)-modes as described in  Eq. \eqref{sim5} we configure the three-point function in such a way that it probes shapes in Fourier space that can be unambiguously connected to the signatures of cosmic string wakes. Consequently, a detection in the context of the three-point function corresponds to a detection of a key string wake signature rather than the detection of some general deviation from the background model.

In the following we focus only on the three-point correlation function. First, we study the string wake detectability as a function  of the string tension \(\mu\). In Table \ref{tab:4} we list the results for different tensions \(\mu\). These results should  be compared with Figure \ref{fig:26}.  We can conclude that Figure \ref{fig:26} provides an accurate description of the detectability of a string wake via the three-point function when the string tension is varied. 
Note that the last entry in Table \ref{tab:4}, the 5th row, describes the result when we include interferometer noise. In our realization of the noise using the baseline density of MWA's Phase I and II we find that around \(5810\) of the \(512\times 512\) pixels corresponding to the smallest \(k\)-modes in Fourier space exhibit a convergent noise level. Consequently, we evaluate the three-point function only on a disc centered at the origin of the angular plane in Fourier space with an area of \(5810\) pixels. In the 5th row of Table \ref{tab:4} we see that the detectability in this case suffers only mildly. This is due to the fact that the instrumental noise is important for larger values of $k$ where the string wake signal is small (see Figure \ref{fig:27}).
\begin{table}[t!]
    \centering
\begin{tabular}{cccccc} \toprule
   row& {angular resolution} &{$G\mu$ [$10^{-6}$]} &  removal factor {$\varepsilon_{fg}$}& detect. \(3\sigma\) & detect. \(5\sigma\) \\ \midrule
   1& \(36''\)&\(0.3\)& \(0.1\)       & + & +\\
   2& \(36''\)&\(0.25\)& \(0.06\)      & + & +\\
   3& \(36''\)&\(0.2\)& \(0.05\)       & + & -\\
   4& \(36''\)&\(0.2\)& \(0.03\)      & + & +\\\toprule
    5& \(36''\)&\(0.3\)& \(0.1\)       & + & -\\
   \bottomrule
\end{tabular}
    \caption{Results of the three-point function \(\braket{T_b(k_1)T_b(k_2)T_b(k_3)}\)-estimator applied on a \(5^\circ\times 5^\circ\) patch in the sky. The configuration of this patch is analogous to the description in Table \ref{tab:3}. Note that in this table we explicitly test at \(z=12\) and apply the Wiener-filter before calculating the three-point function. Also, for the listed results we average over \(10000\) random realizations of the noise contaminants. In the 5th row we include the instrumental noise according to Eq. \eqref{sim1} in the simulated patch based on which we calculate the three-point function. We further elaborate on the effect of this noise component on the detectability of string signatures in the text below.}
    \label{tab:4}
\end{table}
We conclude that the most significant contribution of the string signal resides in lower \(k\)-modes, hence the cut off induced by the instrumental noise for \(u\geq450\) yields merely a slight change in the the detectability. This is shown in Figure (\ref{fig:27}). Note at this point that the angular resolution in Table \ref{tab:4} does not correspond to the angular resolution implied by the antenna configuration in Figure \ref{fig:24}. The former describes the area-to-pixel ratio of the simulated sky patch, while the latter influences the baseline density in the instrumental noise power spectrum and hence restricts the region in \(k\)-space that is not discarded in the calculation of the three-point function due to diverging noise. Hence, even with the chosen area-to-pixel ratio resolution, limitations of the MWA antenna configuration are taken into account. Adjusting the area-to-pixel ratio to the resolution exhibited by the MWA configuration has only negligible impact on the results.

To sum up our results so for: with a residual foreground amplitude reduce by one order of magnitude, a cosmic string wake completely residing in the probed redshift bin is detectable by an interferometer with a noise power spectrum similar to that of the MWA via applying the Wiener-filter and subsequently using the three-point function defined in Eq. \eqref{sim5}.  

\begin{figure}[!tb]
\centering
\includegraphics[scale=0.7]{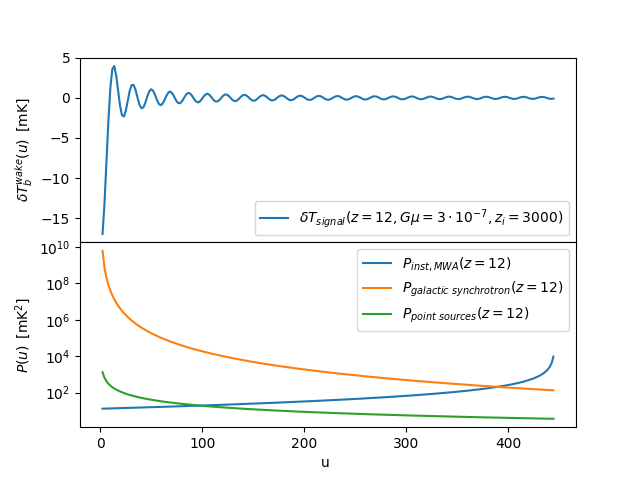}
\caption{In the upper plot we display the string wake signal sliced for \(k_x\) or \(k_y=0\), i.e. we slice through one of the ridges of the wake's signal in Fourier space and display the cross section. The coordinate \(u\) then corresponds to \(u=k_{x/y}/(2\pi)\). \(\delta T_{signal}\) refers to the brightness temperature of the string wake. The lower plot shows the power spectra of the two strongest foreground components and the instrumental noise modeled for MWA's antenna distribution. Note that while the foregrounds are modeled Gaussian in real space, the interferometer noise is modeled as a Gaussian in Fourier space. The coordinate \(u\) corresponds to \(u=l/(2\pi)=\sqrt{k_x^2+k_y^2}/(2\pi)\).}
\label{fig:27}
\end{figure}

To further improve the string wake detectability we now employ the foreground removal scheme sketched in \ref{sec_implement_simulations}.

\subsection{Foreground removal pipeline}

Following the description in section \ref{sec_filters_simulations}, we can simulate multiple consecutive redshift bins enclosing the bin in which the string wake resides in order to interpolate, pixel by pixel, the residual foreground level of the redshift bin containing the string wake. Due to the smooth frequency dependence of the residual foregrounds (see Eq. \eqref{porcozio}), we find a well defined fit function and hence can interpolate very accurately. When applying this technique, we fix the redshift  to \(z=12\), the string tension to \(G\mu=3\cdot 10^{-7}\) and the angular resolution to \(36\) arc-second. Since we aim to mimic a realistic measurement, we include a sinusoidal response by the interferometer to the smooth foregrounds in redshift direction as described in section \ref{sec_implement_simulations}. 

\begin{figure}[!tbp]
\centering
\includegraphics[scale=0.7]{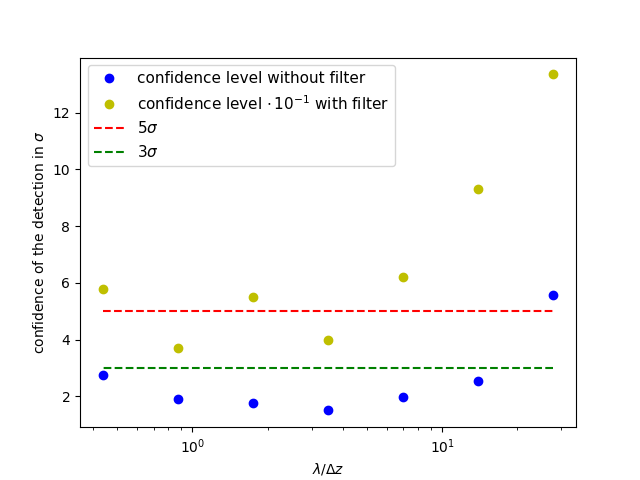}
\caption{Results of Table \ref{tab:5}. The \(3\sigma\) and \(5\sigma\) thresholds are marked with dashed lines. The \(y\)-axis is given in units of \(\sigma\), the \(x\)-axis in units of redshift bin thickness \(\Delta z\). For the data with filter, we divide the confidence level by \(10\) for better visibility and comparability with the case without filter application. The data points display a levelling off for \(\lambda/\Delta z < 2\).}
\label{fig:28}
\end{figure}

The results are presented in Table \ref{tab:5} and Figure \ref{fig:28}. The instrumental response function is modeled via a sine function on top of the foregrounds with an amplitude of one percent of the total foreground amplitude in redshift direction, i.e. it is given by \(0.01\cdot T_b^{fg}(z)\cdot \sin(2\pi/\lambda \cdot (1+z))\). Here, \(\lambda\) is in units of the redshift bin thickness \(\Delta z\), the effective step length in discretized redshift space. We expect the detectability to decrease for lower values of \(\lambda /\Delta z\) as the mode mixing becomes stronger. However, the results indicate a levelling off at low wavelengths. This can be explained by comparing the frequency sampling rate with the wavelength \(\lambda\). In our discrete data set in frequency direction, sampling patches with \(\Delta z \geq \lambda /2\) result in the sine wavelength mimicking a longer wavelength mode as a single redshift bin covers multiple sine function periods. Smaller wavelengths \(\lambda/\Delta z\) on the other hand imply effects similar to just adding a constant to the residual foregrounds. Thus, the detectability function levels off.   

\begin{table}[t!]
    \centering
\begin{tabular}{cccc} \toprule
   row& sine wavelength \(\lambda\) in \([\lambda/\Delta z]\) & detect. with filter [\(\sigma\)] &  detect. without filter [\(\sigma\)]\\ \midrule
   1& \(27.5\)&\( 133.5\)& \(5.58\)   \\
    2& \(13.7\)&\(93.4\)& \(2.6 \)   \\
     3& \(6.9\)&\( 62.8\)& \(2.0\)   \\
      4& \(3.4\)&\(40.9 \)& \( 1.5\)   \\
       5& \(1.7\)&\( 55.1\)& \(1.8 \)   \\
        6& \(0.9\)&\( 37.6\)& \( 1.9\)   \\
        7& \(0.4\)&\( 58.5\)& \(2.8 \)   \\\bottomrule
\end{tabular}
    \caption{Results of the three-point function statistic applied to patches on the sky with dimensions as in previous results (Table \ref{tab:3}, Table \ref{tab:4}). Here, we have applied the foreground removal technique described in \ref{sec_filters_simulations} before calculating the three-point function, and we consider instrumental effects. We fix \(z=12\) and \(G\mu=3\cdot 10^{-7}\). The wavelength \(\lambda\) refers to the wavelength of the sinusoidal response function in redshift direction described in \ref{sec_implement_simulations}. The detectability with or without filter is given in units of \(\sigma\), where we define as above \(\sigma=\sigma_{three-point}/\sqrt{20000/25}\). These results are obtained by averaging over \(1000\)   randomly chosen realizations of the foregrounds with and without the string signal.}
    \label{tab:5}
\end{table}

The results demonstrate that the foreground removal technique works very well and allows us to identify the string signal above all of the noise sources. We can achieve a string detectability at a \(5\sigma\) significance level even without the application of a filter. Note however that this detectability drops below \(3\sigma\) for lower wavelengths of the instrumental response function. The wavelength of the response function is determined by the instrument hardware and its calibration. Here, we have considered several wavelengths to analyse the effectiveness of the removal method for various  potential instrumental and calibration schemes. The results also show that filtering is highly effective.   Wiener-filtering the data after removing the foreground yields a more than \(20\)-fold increase in the significance of the detectability. 

Note that making use of filtering, the string wake is detectable for values of \(\lambda\) in the first two rows with a significance of \(4.7\) to \(3.3\, \sigma_{three-point}\) even if a single  \(5^\circ \times 5^\circ\) patch is being considered. This indicates that if the response function of the interferometer is well understood and its periodicity wavelength is much longer than the considered redshift interval, then the three-point function in combination with foreground removal and filtering is able to detect the string wake with a survey size of a single patch of \(5^\circ \times 5^\circ\) in the sky.  Even if the response function is less favorable for the detection, the large significance for the detectability using filters is promising. The survey size necessary to achieve a \(5\sigma\) detectability scaling sigma as \(\sigma=\sigma_{three-point}/\sqrt{N}\) is much lower than in the case without filter application.

Note that the filters are robust even when the exact orientation of the string wake is unknown.  The degrees of detectability listed in the third column of Table \ref{tab:5} also implies that cosmic strings with a significantly lower tension than the one used as a simulation benchmark can be detected using our data analysis pipeline consisting of foreground removal, filtering and three-point function statistic.

\section{Conclusions and Discussion} \label{conclusion}

We have studied the detectability of a cosmic string wake in interferometric 21-cm surveys. Since a string wake produces a non-Gaussian signal with a distinctive pattern involving edges in Fourier space, a three point function statistic with shape chosen to match the string signal is effective. We have presented the results of mock simulations which include both astrophysical foregrounds and instrumental effects. We find that with an angular resolution of \(36\) arc-seconds, a wake produced by a string with tension $G\mu = 3 \times 10^{-7}$ can be detected with very high significance after the application of Wiener filtering and of the foreground subtraction scheme discussed in this work. The analysis of a single patch of \(5^\circ \times 5^\circ\) in the sky is sufficient to tease out the signal. The analysis of larger patches of the sky will increase the significance of the detection, and will allow the detection of signals from strings with somewhat lower tensions.

Note that we have assumed that our maps contain one wake of angular size \(1^\circ \times 1^\circ\) in the sky produced at a redshift $z_i = 3000$. Based on the scaling solution of the cosmic string network, we know that this is a very conservative assumption. Our sky patch will likely contain a larger number of such string wakes, and this will further improve the detectability. On the other hand, we have assumed optimal orientation of the string for detectability. But since the edge structure in the Fourier maps does not depend on the orientation, our statistic should give similar results also for sub-optimal orientations, and if there are a significant number of string wakes in our patch, then at least one will have close to optimal orientation.

Note that we have modelled our foregrounds as Gaussian processes. This is a rather simplified approximation. However, our main statistic to search for cosmic string signals is a three point function with a specific shape chosen to pick out the planar structures which string wakes create. Since none of the foregrounds are expected to produce such planar structures, we expect the contribution of foregrounds to our three point statistic to be small.

We look forward to applying our analysis scheme to actual data. Another avenue for future research is to explore more sophisticated statistics such as wavelet analyses. It would also be interesting to consider the application of machine learning techniques.

\section*{Acknowledgement}

\noindent This work was funded in part by grant No 200021\_192243 from the Swiss National Science Foundation. RB wishes to thank the Pauli Center and the Institutes of Theoretical Physics and of Particle Physics and Astrophysics of the ETH for hospitality. The research of RB at McGill is supported in part by funds from NSERC and from the Canada Research Chair program. We wish to thank Adrian Liu and Oscar Hernandez for feedback on the draft of this paper.

\section*{Appendix: Wiener and Matched Filtering}

Optimal-filter schemes are commonly used in cosmology and originate from image processing. Two of the most prominent examples known in cosmology are the Wiener-Filter (see e.g. \cite{Wandelt, Bunn} for applications to CMB data ) and the Matched-Filter (see e.g. \cite{Hennawi_2005, Nadathur_2016} for applications to CMB and cluster cosmology ). Both filters are similar in their implementation. They rely on the approximate knowledge of the power spectrum of the signal we want to extract and of the noise contamination in the analyzed data. With these power spectra we can construct a function in Fourier space that can be multiplied with the data so that it enhances the string wake signal while simultaneously damping the noise contamination.

The filters are modeled as follows. At a fixed frequency, we assume our incoming signal to have the following form
\begin{align}
    f(x,y) \, = \, s(x,y) + n(x,y).
\end{align}
Here, \(f\) represents the received data, \(s\) represents the signal (in our case the wake signal) and \(n\) the noise, i.e. the residual foregrounds. The coordinates \((x,y)\) parametrize the two-dimensional real space. We aim to determine a function \(g(x,y)\) so that 
\begin{align}\label{sim6}
    \Tilde{f}(x,y) \, = \, (g\star f)(x,y)
\end{align}
is an estimator for \(f\) minimizing the mean square error and \(\star\) represents the mathematical convolution. The advantage of this formulation is that in Fourier space the convolution boils down to the product of the Fourier transforms. 

For the Wiener-filter the function \(g\) is defined via its Fourier transform
\begin{align}
    G(k_x,k_y) \, = \, \frac{S(k_x,k_y)}{S(k_x,k_y) + N(k_x,k_y)},
\end{align}
where \(S\) and \(N\) are the power spectra of the signal and noise and \(k_x,k_y\) are the coordinates in Fourier space. For the Matched-Filter, it is defined as 
\begin{align}
    G(k_x,k_y) \, = \, \frac{S(k_x,k_y)}{N(k_x,k_y)}.
\end{align}
If we have a well described signal and noise, i.e. if the power spectra are known, we can calculate and insert the functions \(G\) into the Fourier transform of Eq. \eqref{sim6}. The resulting function 
\begin{align}
    \Tilde{F}(k_x,k_y) \, = \, G(k_x,k_y) \cdot F(k_x,k_y)
\end{align}
can be inverse Fourier transformed. In doing so, we recover a function describing the incoming signal of the original measurement but with the target signature being more pronounced with respect to the noise contamination.
 
The Wiener- and Matched-filters are processing techniques that are comparably easy to implement. Nonetheless, they have certain advantages for the analysis of signatures in Fourier space. Both filters preserve the phase information of the data \(f\). Due to \(G\) being purely real as a combination of real power spectra\footnote{We define the power spectrum \(P\) as \(P(k)\sim |F(k_x,k_y)|^2\).}, the imaginary parts of \(F\) and \(\Tilde{F}\)  agree. Additionally, after application of the filters the residual noise component left in the data is approximately white noise, i.e. the covariance matrix in real space is diagonal. This significantly eases the analysis in real space. Note, however, that since interferometer data is sampled in Fourier space and constructing real space maps based on this data generally comes with side effects that may impair the string wake detectability, it is reasonable to conduct the whole analysis in Fourier space as it is done in the context of this work.

Wiener- and Matched-filters are generally robust. We use robust here in the sense that even if the theoretical description of the signal or noise power spectrum is not in perfect alignment with the real power spectrum of the corresponding component, the filters still improve the signal to noise ratio of the data. This is a huge advantage when searching for string wakes. The shape of the wake power spectrum generally depends on the alignment of the wake within the redshift bin (or across many redshift bins). Selecting a reference alignment, the filters are capable of picking up signals that exhibit small deviations with respect to the signature shape of the reference alignments. Thus, we can probe for multiple string alignments at once. As the wake alignment is in general unknown, a practical approach for the application of the filter in the context of string detection would be to filter the measurement data with the Wiener- or the Matched-filter for multiple distinct wake power spectra.  

All in all, filtering techniques are a good first approach for increasing the signal to noise ratio. In particular, their robustness in the context of approximate theoretical descriptions for the power spectra makes them suitable for the search for cosmic string signatures. Note however, that the filters are applied to data of two-dimensional patch and do not include information from the third dimension of interferometer data, the frequency direction. In the following, we aim to use our knowledge of the extent of the residual foregrounds in redshift space to further improve the detectability of the cosmic string wake.

\end{document}